\documentclass[prd,preprintnumbers,pacs,12pt]{revtex4}
\usepackage{epsfig}
\usepackage{graphicx}
\usepackage {axodraw}

\usepackage{bm}
\usepackage{amsmath}
\usepackage{amssymb}
\usepackage{amscd}
\usepackage{latexsym}

\newcommand{\be}{\begin{equation}}

\newcommand{\ee}{\end{equation}}
\newcommand{\bea}{\begin{eqnarray}}
\newcommand{\eea}{\end{eqnarray}}

\newcommand{\nn}{\nonumber}

\def\Black{}
 \def\AliasBlue{}
 \def\Blue{}
 \def\Brown{}

\begin{document}

\newcommand{\bra}[1]{\langle #1|}
\newcommand{\ket}[1]{|#1\rangle}
\newcommand{\braket}[2]{\langle #1|#2\rangle}
\newcommand{\tr}{\textrm{Tr}}
\newcommand{\lag}{\mathcal{L}}
\newcommand{\mbf}[1]{\mathbf{#1}}
\newcommand{\desl}{\slashed{\partial}}
\newcommand{\Desl}{\slashed{D}}
%\preprint{{\bf DFF-413/05/04}}

\renewcommand{\bottomfraction}{0.7}
\newcommand{\epsi}{\varepsilon}

\newcommand{\nl}{\nonumber \\}
\newcommand{\tc}[1]{\textcolor{#1}}
\newcommand{\sla}{\not \!}
\newcommand{\spinor}[1]{\left< #1 \right>}
\newcommand{\cspinor}[1]{\left< #1 \right>^*}
\newcommand{\Log}[1]{\log \left( #1\right) }
\newcommand{\Logq}[1]{\log^2 \left( #1\right) }
\newcommand{\mr}[1]{\mathrm{#1}}
\newcommand{\cw}{c_\mathrm{w}}
\newcommand{\sw}{s_\mathrm{w}}
\renewcommand{\i}{\mathrm{i}}
\renewcommand{\Re}{\mathrm{Re}}
\newcommand{\yText}[3]{\rText(#1,#2)[][l]{#3}}
\newcommand{\xText}[3]{\put(#1,#2){#3}}

\title{Z$^\prime$ production at the LHC in the four-site Higgsless model}

\author{Elena Accomando$^{(a)}$, Stefania De Curtis$^{(b)}$, Daniele Dominici$^{(b,c)}$ and Luca Fedeli$^{(b,c)}$}

\affiliation
{$(a)$ 
NExT Institute and School of Physics and Astronomy, University of Southampton, Highfield,
Southampton SO17 1BJ, UK\\
$(b)$ INFN, 50019 Sesto F., Firenze, Italy\\
 $(c)$ Department of Physics, University of Florence, 50019 Sesto F., Firenze, Italy.}

%\date{\today}

\begin{abstract}
We study the phenomenology of the neutral gauge sector of the four-site 
Higgsless model, based on the $SU(2)_L\times SU(2)_1\times SU(2)_2\times 
U(1)_Y$ gauge symmetry, at present colliders. The model predicts the 
existence of two neutral and four charged extra gauge bosons, $Z_{1,2}$ and 
$W^\pm_{1,2}$. We expand and update a previous study, by concentrating on the 
neutral sector. We derive new limits on $Z_{1,2}$-boson masses and 
couplings from recent direct searches at the Tevatron. We moreover estimate 
the discovery potential at the Tevatron with a project luminosity 
L=10 fb$^{-1}$, and at the 7 TeV LHC with L=1 fb$^{-1}$.

In contrast to other Higgsless theories characterized by almost fermiophobic 
extra gauge bosons, the four-site model allows sizeable $Z_{1,2}$-boson 
couplings to SM fermions. Owing to this feature, we find that in the next two 
years the extra $Z_{1,2}$-bosons could be discovered in the favoured 
Drell-Yan channel at the 7 TeV LHC for $Z_{1,2}$ masses in 
the TeV region, depending on model parameters.

\end{abstract}

\maketitle

\def\to{\rightarrow}
\def\ptl{\partial}
\def\beq{\begin{equation}}
\def\eeq{\end{equation}}
\def\bea{\begin{eqnarray}}
\def\eea{\end{eqnarray}}
\def\nn{\nonumber}
\def\half{{1\over 2}}
\def\rhalf{{1\over \sqrt 2}}
\def\calo{{\cal O}}
\def\call{{\cal L}}
\def\calm{{\cal M}}
\def\del{\delta}
\def\eps{\epsilon}
\def\lam{\lambda}

\def\anti{\overline}
\def\delfac{\sqrt{{2(\del-1)\over 3(\del+2)}}}
\def\heff{h'}
\def\square{\boxxit{0.4pt}{\fillboxx{7pt}{7pt}}\hspace*{1pt}}
    \def\boxxit#1#2{\vbox{\hrule height #1 \hbox {\vrule width #1
             \vbox{#2}\vrule width #1 }\hrule height #1 } }
    \def\fillboxx#1#2{\hbox to #1{\vbox to #2{\vfil}\hfil}   }

\def\braket#1#2{\langle #1| #2\rangle}
\def\gev{~{\rm GeV}}
\def\gam{\gamma}
\def\sn{s_{\vec n}}
\def\sm{s_{\vec m}}
\def\mm{m_{\vec m}}
\def\mn{m_{\vec n}}
\def\mh{m_h}
\def\sumn{\sum_{\vec n>0}}
\def\summ{\sum_{\vec m>0}}
\def\vl{\vec l}
\def\vk{\vec k}
\def\ml{m_{\vl}}
\def\mk{m_{\vk}}
\def\gp{g'}
\def\gt{\tilde g}
\def\hw{{\hat W}}
\def\hz{{\hat Z}}
\def\ha{{\hat A}}

\def\yy{{\cal Y}_\mu}
\def\yyt{{\tilde{\cal Y}}_\mu}
\def\lq{\left [}
\def\rq{\right ]}
\def\dmu{\partial_\mu}
\def\dnu{\partial_\nu}
\def\dmus{\partial^\mu}
\def\dnus{\partial^\nu}
\def\gp{g'}
\def\gpt{{\tilde g'}}
\def\gs{g''}
\def\ggs{\frac{g}{\gs}}
\def\eps{{\epsilon}}
\def\tr{{\rm {tr}}}
\def\V{{\bf{V}}}
\def\W{{\bf{W}}}
\def\Wt{\tilde{ {W}}}
\def\Y{{\bf{Y}}}
\def\Yt{\tilde{ {Y}}}
\def\L{{\cal L}}
\def\s{s_\theta}
\def\st{s_{\tilde\theta}}
\def\c{c_\theta}
\def\ct{c_{\tilde\theta}}
\def\gt{\tilde g}
\def\et{\tilde e}
\def\At{\tilde A}
\def\Zt{\tilde Z}
\def\Wpt{{\tilde W}^+}
\def\Wmt{{\tilde W}^-}

\newcommand{\Apt}{{\tilde A}_1^+}
\newcommand{\Bpt}{{\tilde A}_2^+}
\newcommand{\Amt}{{\tilde A}_1^-}
\newcommand{\Bmt}{{\tilde A}_2^-}
\newcommand{\Wtp}{{\tilde W}^+}
\newcommand{\Atp}{{\tilde A}_1^+}
\newcommand{\Btp}{{\tilde A}_2^+}
\newcommand{\Atm}{{\tilde A}_1^-}
\newcommand{\Btm}{{\tilde A}_2^-}
\def\mathswitchr#1{\relax\ifmmode{\mathrm{#1}}\else$\mathrm{#1}$\fi}
\newcommand{\Pe}{\mathswitchr e}
\newcommand{\Pp}{\mathswitchr {p}}
\newcommand{\PZ}{\mathswitchr Z}
\newcommand{\PW}{\mathswitchr W}
\newcommand{\PD}{\mathswitchr D}
\newcommand{\PU}{\mathswitchr U}
\newcommand{\PQ}{\mathswitchr Q}
\newcommand{\Pd}{\mathswitchr d}
\newcommand{\Pu}{\mathswitchr u}
\newcommand{\Ps}{\mathswitchr s}
\newcommand{\Pc}{\mathswitchr c}
\newcommand{\Pt}{\mathswitchr t}
\newcommand{\rd}{{\mathrm{d}}}
\newcommand{\GW}{\Gamma_{\PW}}
\newcommand{\GZ}{\Gamma_{\PZ}}
\newcommand{\GeV}{\unskip\,\mathrm{GeV}}
\newcommand{\MeV}{\unskip\,\mathrm{MeV}}
\newcommand{\TeV}{\unskip\,\mathrm{TeV}}
\newcommand{\fba}{\unskip\,\mathrm{fb}}
\newcommand{\pba}{\unskip\,\mathrm{pb}}
\newcommand{\nba}{\unskip\,\mathrm{nb}}
\newcommand{\PT}{P_{\mathrm{T}}}
\newcommand{\PTmiss}{P_{\mathrm{T}}^{\mathrm{miss}}}
\newcommand{\CM}{\mathrm{CM}}
\newcommand{\inv}{\mathrm{inv}}
\newcommand{\sig}{\mathrm{sig}}
\newcommand{\tot}{\mathrm{tot}}
\newcommand{\backg}{\mathrm{backg}}
\newcommand{\evt}{\mathrm{evt}}
% particle masses
\def\mathswitch#1{\relax\ifmmode#1\else$#1$\fi}
\newcommand{\M}{\mathswitch {M}}
\newcommand{\R}{\mathswitch {R}}
\newcommand{\TEV}{\mathswitch {TEV}}
\newcommand{\LHC}{\mathswitch {LHC}}
\newcommand{\MW}{\mathswitch {M_\PW}}
\newcommand{\MZ}{\mathswitch {M_\PZ}}
\newcommand{\Mt}{\mathswitch {M_\Pt}}
\def\si{\sigma}
\def\beqar{\begin{eqnarray}}
\def\eeqar{\end{eqnarray}}
\def\refeq#1{\mbox{(\ref{#1})}}
\def\reffi#1{\mbox{Fig.~\ref{#1}}}
\def\reffis#1{\mbox{Figs.~\ref{#1}}}
\def\refta#1{\mbox{Table~\ref{#1}}}
\def\reftas#1{\mbox{Tables~\ref{#1}}}
\def\refse#1{\mbox{Sect.~\ref{#1}}}
\def\refses#1{\mbox{Sects.~\ref{#1}}}
\def\refapps#1{\mbox{Apps.~\ref{#1}}}
\def\refapp#1{\mbox{App.~\ref{#1}}}
\def\citere#1{\mbox{Ref.~\cite{#1}}}
\def\citeres#1{\mbox{Refs.~\cite{#1}}}

\def\Black{}
 \def\AliasBlue{}
 \def\Blue{}
 \def\Brown{}

%\section{Introduction}

%\section{Review of the model}
%\label{linear}

%\subsection{Fermion gauge boson couplings}

%\subsection{Perturbative unitarity bounds}

\section{Introduction}

Models with additional extra-dimensions 
\cite{ArkaniHamed:1998rs,Antoniadis:1998ig,Randall:1999vf,Randall:1999ee} 
have received, during the last ten years, considerable attention because they 
are a new interesting attempt to explain the large difference between the 
Planck scale and the Fermi scale via geometrical arguments. They also provide 
new mechanisms of electroweak symmetry breaking (EWSB). The symmetry could in 
fact be broken either by orbifolding or via boundary conditions on the 
compact extra dimensions. As a consequence, a remarkable activity has been 
devoted to investigate Higgsless  Models (HM) 
\cite{Csaki:2003dt,Csaki:2003zu,Barbieri:2003pr,Cacciapaglia:2004zv,
Cacciapaglia:2004jz}. HM, being 5D gauge models, predict the existence of new 
Kaluza-Klein (KK) vector resonances, which help in delaying the unitarity 
violation of vector boson scattering (VBS) amplitudes to higher energies, 
compared to the corresponding scale of the Standard Model (SM) without a 
light Higgs \cite{SekharChivukula:2001hz,Csaki:2003dt}. The discretization of 
the compact fifth dimension, over which HM are defined, generates the so-called
deconstructed theories which are described by 4D chiral lagrangians with a 
number of replicas of the gauge group equal to the number of lattice sites
 \cite{ArkaniHamed:2001ca,Arkani-Hamed:2001nc,Hill:2000mu,Cheng:2001vd,
Abe:2002rj,Falkowski:2002cm,Randall:2002qr,Son:2003et,deBlas:2006fz}. 
The main difficulty for all these models, as for technicolor theories, is to 
reconcile the presence of a relatively low KK-spectrum, necessary to delay 
the unitarity violation to TeV-energies, with the electroweak precision tests 
(EWPT) whose measurements are in very good agreement with SM predictions. 
One possible solution is obtained by either delocalizing fermions along the 
fifth dimension \cite{Cacciapaglia:2004rb,Foadi:2004ps} or, equivalently in 
the deconstructed picture, by allowing for direct couplings between new 
vector bosons and SM fermions \cite{Casalbuoni:2005rs}. In the simplest 
version of this latter class of models, corresponding to just three lattice
sites and gauge symmetry $SU(2)_L\times SU(2)\times U(1)_Y$ (the so-called 
BESS model \cite{Casalbuoni:1985kq,Casalbuoni:1986vq}), the requirement of
vanishing of the $\eps_3 (S)$ parameter implies that the new triplet of
vector bosons is almost fermiophobic; then the only production channels for 
their search are those driven by boson-boson couplings. The HM literature has 
been thus mostly focused on difficult multi-particle processes which require 
high luminosity to be detected, that is vector boson fusion and associated 
production of new gauge bosons with SM ones 
\cite{Birkedal:2004au,He:2007ge,Belyaev:2007ss,Alves:2008up,Alves:2009aa}. 

In a recent paper \cite{Accomando:2008jh}, we have considered the 
generalization of the minimal three-site model by inserting an additional 
lattice site. The four-site
Higgsless model, which emerges, is based on the $SU(2)_L\times SU(2)_1\times
SU(2)_2\times U(1)_Y$ gauge symmetry and predicts two neutral and four
charged extra gauge bosons, $Z_{1,2}$ and $W^\pm_{1,2}$. Its novelty and 
strenght consist in satisfying EWPT constraints without imposing the new 
resonances to be fermiophobic. Within this framework, the favoured Drell-Yan 
channel becomes particularly relevant for the extra gauge boson search at the 
LHC \cite{Accomando:2008dm,Accomando:2008jh,DeCurtis:2010sz} and the upgraded
Tevatron.

Aim of this paper is to update the phenomenological aspects of the KK neutral 
sector, studied in \cite{Accomando:2008dm,Accomando:2008jh}, by taking into 
account recent data from Tevatron and the new plans of the 7 TeV LHC.
We compare the sensitivity reach of the Tevatron with 10 fb$^{-1}$ of 
integrated luminosity and the LHC at 7 TeV with L=1 fb$^{-1}$. We also study 
how the sensitivity reach could improve at the 14 TeV LHC with L=10 fb$^{-1}$.

The paper is organized as follows. In Section \ref{EWPT_UNIT}, we briefly 
review the model. In Section \ref{spectrum}, we discuss mass spectrum, decay 
widths and branching ratios of the two additional neutral bosons. In Section 
\ref{dy}, we perform a detailed analysis of $Z_{1,2}$ Drell-Yan 
cross-sections and invariant mass distributions. We compare the results 
expected in the next two years at the Tevatron with L=10 fb$^{-1}$ and the 7 
TeV LHC with L=1 fb$^{-1}$, with those ones reachable at the upgraded 14 TeV 
LHC with L=10 fb$^{-1}$. In Section \ref{disco}, we derive exclusion limits 
on the $Z_{1,2}$ bosons from Tevatron and LHC experiments, and we discuss 
their discovery prospects. Conclusions are given in Section 
\ref{conclusions}. In Appendix \ref{appendixA}, we list all relevant 
trilinear gauge boson couplings.

\section{The model: Unitarity and EWPT bounds}\label{EWPT_UNIT}

The class of models, we are interested in, is described by a chiral Lagrangian based on the $SU(2)_L\otimes SU(2)^K\otimes U(1)_Y$ gauge symmetry and $K+1$ 
non linear $\sigma$-model scalar fields $\Sigma_i$, interacting with the 
gauge fields. The $K$ gauge coupling constants, $g_i$, and the $K+1$ link 
couplings, $f_i$, define the free parameters of the model beyond the SM ones. 
Different $f_i$, in the continuum limit, can describe a generic  metric 
\cite{Son:2003et,Foadi:2003xa,Casalbuoni:2004id,Foadi:2004ps,
Casalbuoni:2005rs}. The flat metric corresponds to constant link couplings: 
$f_i=f_c (i=1,K+1)$. $SU(2)_L$ and $U(1)_Y$ gauge symmetries act on the two 
ends of the linear moose, which in the continuum become the boundaries of the 
fifth dimension. Standard fermions are localized on those ends and do not 
interact with the internal gauge bosons. However, direct couplings between SM 
fermions and new gauge bosons can be included in a way that preserves the 
symmetry of the model. This can be achieved either by considering composite 
operators \cite{Casalbuoni:1985kq,Casalbuoni:1986vq,Casalbuoni:2005rs} or via 
a mixing with new heavy fermions interacting with the new gauge bosons 
\cite{SekharChivukula:2006cg}. We consider only direct couplings between new 
gauge bosons and left-handed SM fermions, with strength given by the $K$ 
dimensionless parameters $b_i$. The case $K=1$ corresponds to the BESS 
model \cite{Casalbuoni:1985kq,Casalbuoni:1986vq} with a particular choice of 
the parameters. Here, we concentrate on the case $K=2$ which in turn 
corresponds to a specific choice of the parameters of the vector and axial 
vector-extension of the BESS model \cite{Casalbuoni:1989xm}. For simplicity, 
we assume $g_2=g_1$ and $f_3=f_1$. This choice leads to a Left-Right symmetry
in the new gauge sector, giving rise to a definite parity for the 
corresponding gauge bosons once the standard gauge interactions are turned 
off. Summarizing, the four-site Higgsless model has five parameters
a priori: $g_1$, $f_1$, $f_2$, $b_1$ and $b_2$. However, this number gets 
reduced to three by phenomenological constraints as discussed in the following.

Charged and neutral gauge boson spectrum and couplings to SM fermions are 
given in \cite{Accomando:2008jh,Accomando:2008dm} by means of a perturbative 
expansion in series of $x^2=\tilde g^2/g_1^2$. Neutral non zero mass 
eigenvalues have the following expressions, up to ${\mathcal O}(x^4)$:
\be
 M_Z^2=\tilde{M}_Z^2\left(1-
x^2 z_Z\right) 
\ee 
\be \label{A14}
M_{Z_1}^2=M_1^2\left(1+ \frac{ x^2}{2c^2_{\tilde{\theta}} }\right) 
\ee
\be 
M_{Z_2}^2=M_2^2 \left(1+ \frac{ x^2z^2}{2c^2_{\tilde{\theta}} }\right) \label{eq:M2}
\ee
 where 
\be\label{A15}
\tilde{M}_Z^2=\frac{\tilde{M}_W^2}{c^2_{\tilde{\theta}}},\quad \tilde{M}_W^2=\tilde g^2v^2/4,
~~~~~z_Z=\frac{1}{2}\frac{(z^4+c^2_{2\tilde{\theta}})}{c^2_{\tilde{\theta}}}
\ee 
\be\label{A16}
M_1^2=f_1^2 g_1^2, \quad M_2^2=g_1^2(f_1^2 + 2 f_2^2),\quad z=M_1/M_2<1\quad \mbox{and}\quad
t_{\tilde{\theta}} =
s_{\tilde{\theta}}/c_{\tilde{\theta}}
\ee

In the mass eigenstate basis, the Lagrangian describing the neutral 
fermion-boson interaction is given by:
%\footnote{For details see \cite{Accomando:2008jh}.}
\begin{equation}\label{NC}
{\mathcal L}_{NC}=\bar{\psi} \gamma^\mu \left [- e \mathbf{Q}_f A_\mu 
+ ({a}_{ZL}^fP_L+{a}_{ZR}^fP_R) Z_\mu
+({a}_{1L}^fP_L+{a}_{1R}^fP_R)
Z_{1\mu}+ ({a}_{2L}^fP_L+{a}_{2R}^fP_R) Z_{2\mu}  \right ]\psi
\end{equation}
where  $P_{L,R}=(1\mp \gamma_5)/2$ and ${a}_{ZL(R)}^f $, ${a}_{1,2L(R)}^f$ 
are the left-handed (right-handed) couplings of $Z$, $Z_{1,2}$ bosons to SM 
fermions respectively.
Their explicit expressions are given by \cite{Accomando:2008jh}
\be  a_{ZL}^f=-\frac{\gt}{c_ {\tilde{\theta}}}
\left(1-\frac{b}{2}\right)\left(1-x^2\frac{z_Z}{2}\right)\left[\frac{\tau^{f}_3}2-\frac{s^2_
{\tilde{\theta}}}{1-\frac{b}{2}}\left(1-x^2 \frac{ c_
{\tilde{\theta}}}{s_
{\tilde{\theta}}}z_{Z\gamma}\right)Q^f\right] \ee

\be  a_{ZR}^f=\frac{\gt}{c_ {\tilde{\theta}}}
\left(1-\frac{b}{2}\right)\left(1-x^2\frac{z_Z}{2}\right)\left[\frac{s^2_
{\tilde{\theta}}}{1-\frac{b}{2}}\left(1-x^2 \frac{c_
{\tilde{\theta}}}{s_
{\tilde{\theta}}}z_{Z\gamma}\right)Q^f\right] \ee
\be
{a}^f_{1L} =-\frac{\gt x }{\sqrt{2}(1+b_+)}\left(\frac{b_+}{x^2}
- \frac{c_{ 2\tilde{\theta}}}{c^2
_{\tilde{\theta}}}\right)\frac {\tau_3^f} 2 +\frac{\gt xt^2_{\tilde{\theta}}}{\sqrt{2} } Q^f \ee 
\be
{a}^f_{1R} =\frac{\gt xt^2_{\tilde{\theta}}}{\sqrt{2} } Q^f \ee

\be\label{a2L}
{a}_{2L}^f=-\frac{\gt x}{\sqrt{2}(1+b_+)}\left(\frac{b_-}{x^2}
-\frac{z^2}{c^2 _{\tilde{\theta}}}
\right)\frac{\tau_3^f} 2-\frac{ \gt x z^2
t^2_{\tilde{\theta}}}{\sqrt{2} }  Q^f \ee 
\be\label{a2R}
{a}_{2R}^f=-\frac{ \gt x z^2
t^2_{\tilde{\theta}}}{\sqrt{2} }  Q^f \ee 
where
\be
z_{Z\gamma}=-t_{\tilde{\theta}}c_{ 2\tilde{\theta}} \quad\quad
 b=\frac{b_+-b_-z^2}{(1 +b_+) }
\quad\quad b_\pm =b_1\pm b_2 \quad\quad
\ee 
${Q^f}$ is
the electric charge in unit $e$ (the proton charge).
In the following we will factorize out the electric charge by defining
$a^f_{iL,R}=- e \hat{a}^f_{iL,R}$, $i=1,2$.

Using these equations, we can rewrite the free parameters of the four-site 
model in terms of observable quantities. The physical $Z_{1,2}$-boson masses,
$M_{Z_1,Z_2}$, are indeed closely related to $f_{1,2}$ as shown in 
Eqs.~(\ref{A14}-\ref{A16}). Moreover, the $Z_{1,2}$-boson couplings to SM 
fermions, $\hat{a}^f_{1L}$ and $\hat{a}^f_{2L}$, are unique functions of 
$b_{1,2}$ once $g_1,\; M_{Z_1,Z_2}$ are fixed. We can then recast the free 
parameter set $(g_1, f_1, f_2, b_1, b_2)$ previously given into
$(g_1, M_{Z_1}, M_{Z_2}, \hat{a}^\Pe_{1L}, \hat{a}^\Pe_{2L})$.
Having everything expressed in terms of physical quantities, we are now ready 
to discuss the phenomenological constraints. 
   
We get a first relation among the newly defined parameters of the model by 
imposing the SM gauge boson masses to have the measured values. Namely we get at the leading order:
\be\label{x-parameter}
x\sim\frac{M_W}{M_1} \sqrt{\frac{2}{1-z^2}}\sim\frac{c_{\tilde\theta} M_Z}{M_1}\sqrt{\frac{2}{1-z^2}}
\ee
This reduces the number of independent parameters by unit
as $g_1=g_1(M_{Z_1}, M_{Z_2}, \hat{a}^f_{1L}, \hat{a}^f_{2L})$.

\begin{figure}[!htbp]
\begin{center}
\includegraphics[width=6. cm]{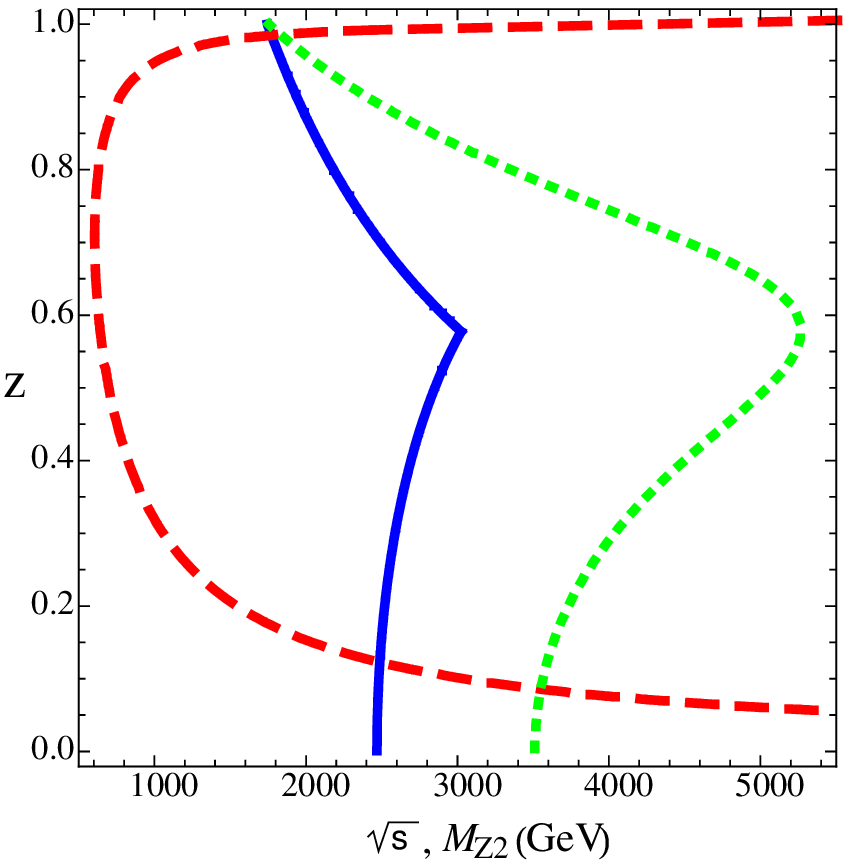}~~~~~~~~~
\includegraphics[width=6.3 cm]{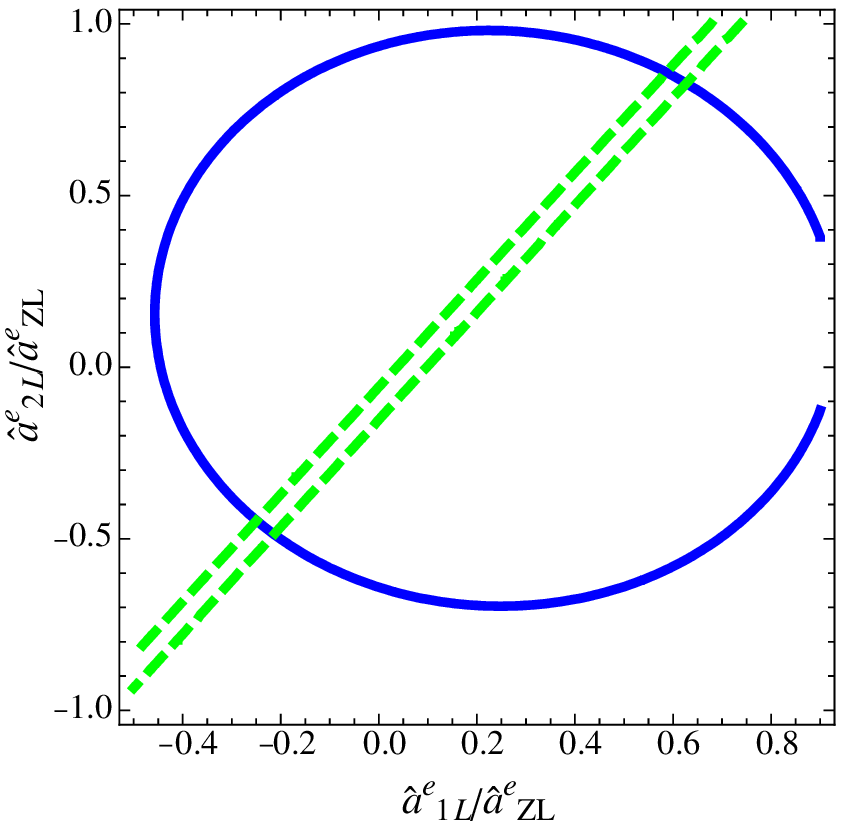}
\end{center}
\caption{Left: Unitarity bounds on the plane ($\sqrt{s}, z$) coming from Vector Boson Scattering (VBS) 
amplitudes with purely SM bosons in the external legs (green dot line), and 
with SM plus extra gauge bosons in the external legs (blue solid line). The 
allowed regions are on the left of the contours. 
The red dashed line gives the minimum value of the $Z_2$-boson
mass spectrum (and accordingly the minimum value of $M_{Z_1}$ which can be 
derived from $z$) allowed by the four-site model at leading order (see text).
Right: 95\% C.L. bounds on the plane 
$(\hat{a}_{1L}^\Pe/\hat{a}_{ZL}^\Pe,\hat{a}_{2L}^\Pe/\hat{a}_{ZL}^\Pe)$ from 
$\epsilon_3$ (green dashed line) and $\epsilon_1$ (blue solid line) for 
$z=0.8$ and $M_1=1$ TeV (corresponding to $M_{Z_{1},Z_2}=1012,\;1256\quad\!\!\!\!\!{\rm GeV}$). The allowed regions are 
the internal ones.}
\label{bi1}
\end{figure}

In addition, the abovementioned relation has an important consequence on the 
allowed spectrum for the new $Z_{1,2}$ gauge bosons. Since mass eigenvalues 
and eigenvectors have been computed via an analytical expansion in the 
$x$-variable, in Fig.~\ref{bi1} (left panel) we plot the validity range of 
this 
approximation. The dashed-line contour corresponds to $x^4=5\times 10^{-3}$, 
and defines the minimum $Z_2$-boson mass allowed by the model (at leading 
order) as a function of the free parameter $z$. Hence, being $z$ the ratio 
between $Z_1$ and $Z_2$-boson masses, a lower bound on the global mass 
spectrum is established. In the following we will refer to this bound as the approximation limit.
 Next we remind that, owing to the exchange of the 
extra gauge bosons, the perturbative unitarity violation of the longitudinal 
vector boson scattering amplitudes with purely SM external bosons (including 
all external gauge bosons) can be delayed up to $\sqrt{s}=5(3)$TeV
\cite{Accomando:2008jh,Accomando:2008dm}. This condition gives an upper bound
on the mass spectrum of the new particles, which are thus constrained to be 
between few hundreds GeV and few TeV as displayed in Fig.~\ref{bi1} (left 
panel). 

In general, the only way to combine the need of relatively light extra gauge
bosons with EWPT is to impose the new particles to be fermiophobic. In the 
four-site Higgsless model, this strong assumption is not necessary anymore. 
In order to show this feature, we compute the new physics contribution to the 
electroweak parameters $\epsilon_1$, $\epsilon_2$ and $\epsilon_3$, including 
one-loop radiative corrections (they are evaluated within the 
Higgsless SM for a 1 TeV cutoff). Then we compare the result with the 
$\epsilon_i$ (i=1,3) experimental values \cite{Barbieri:2004qk}. For fixed 
$Z_{1,2}$-boson masses, one obtains bounds on the two remaining free 
parameters, which we choose to 
be $\hat{a}^\Pe_{1L}$ and $\hat{a}^\Pe_{2L}$, the $Z_{1,2}-$boson couplings 
to the left-handed electron.  
The result is shown in Fig.~\ref{bi1} (right panel) for the sample case: 
$z=0.8$ and $M_1=1$ TeV corresponding to ($M_{Z_1,Z_2}=1012,\;1256\quad\!\!\!\!\!{\rm GeV}$). 
%Here the bound from $\epsilon_3$ goes from up to down by increasing $M_2$,
%that from $\epsilon_1$ is quite insensitive to the value of the
%resonance masses 
No bound comes from $\epsilon_2$ thanks to its negative experimental value.
The $\epsilon_1$ parameter gives weak limits on the magnitude of the $Z_{1,2}$ 
boson-electron couplings. As a consequence, they are free to be of the order 
of the SM ones. This clearly shows that, oppositely to most common Higgsless 
theories present in the literature, the four-site model is not-fermiophobic 
at all and could be proved in the favoured Drell-Yan channel already at the 
LHC start-up. 

\begin{figure}[h!]
\begin{center}
\unitlength1.0cm
\begin{picture}(7,6)
\put(-3.7,0.4){\epsfig{file=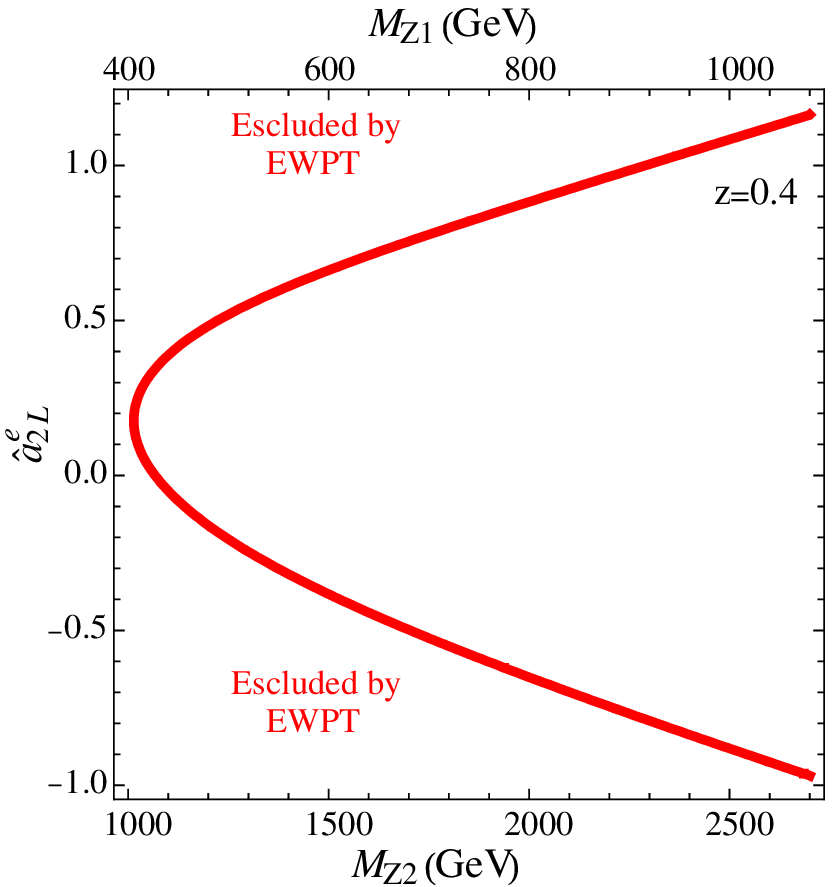,width=5.5cm}}
%\put(-3.7,1.8){\epsfig{file=z04_EWPT.eps,width=5.5cm}}
\put(3.9,0.4){\epsfig{file=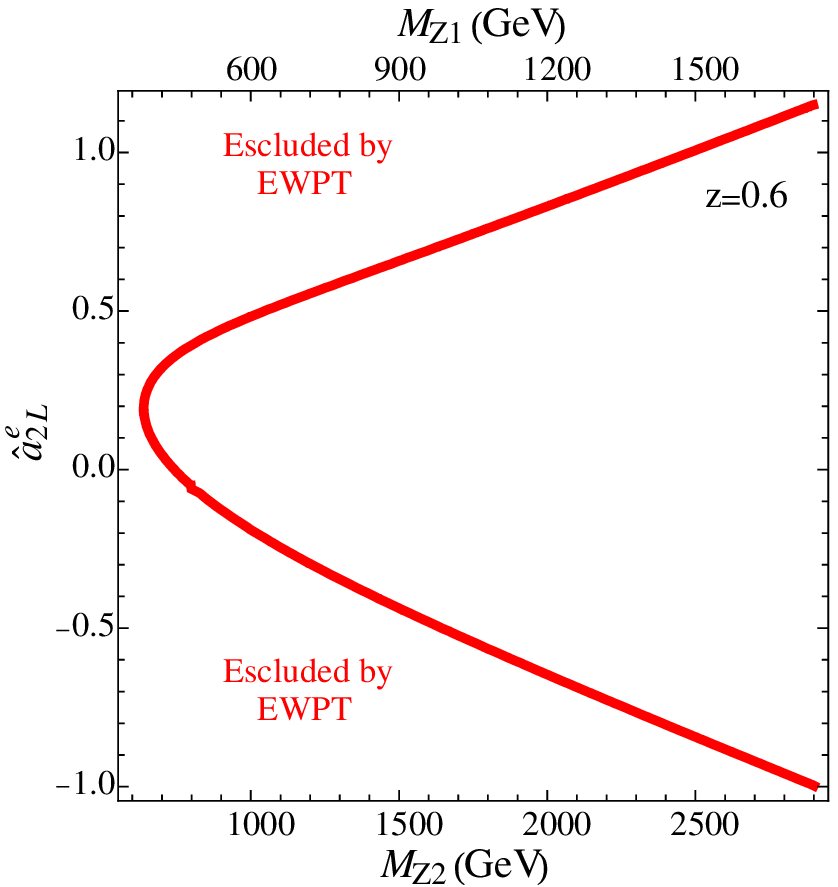,width=5.5cm}}
\put(-3.7,-6.3){\epsfig{file=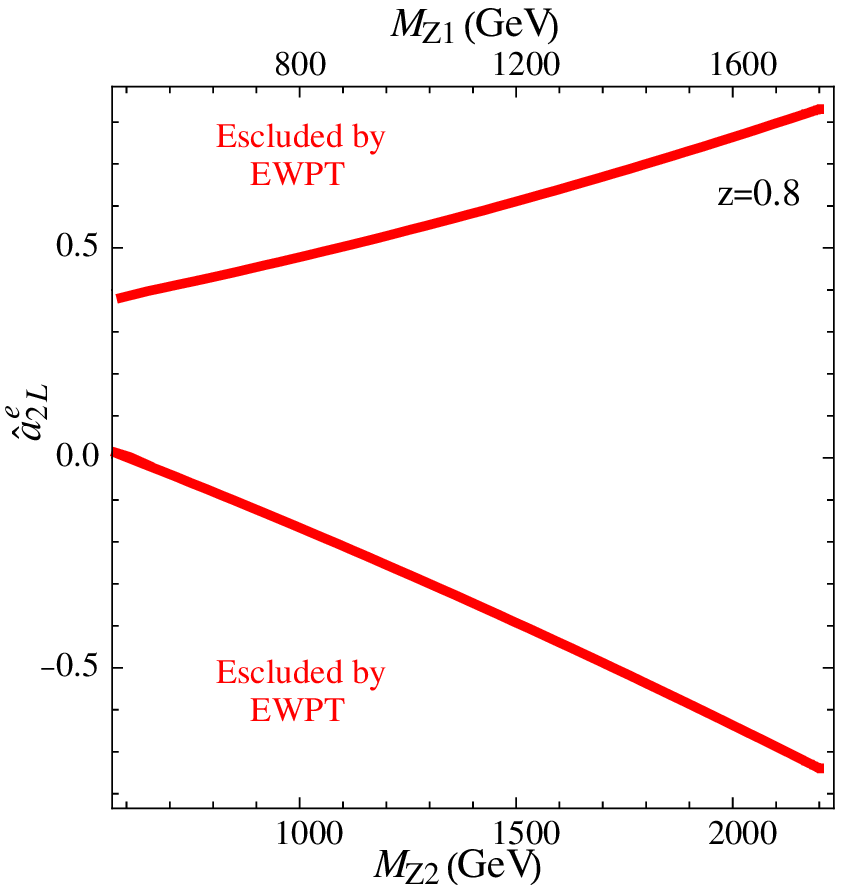,width=5.5cm}}
\put(3.9,-6.3){\epsfig{file=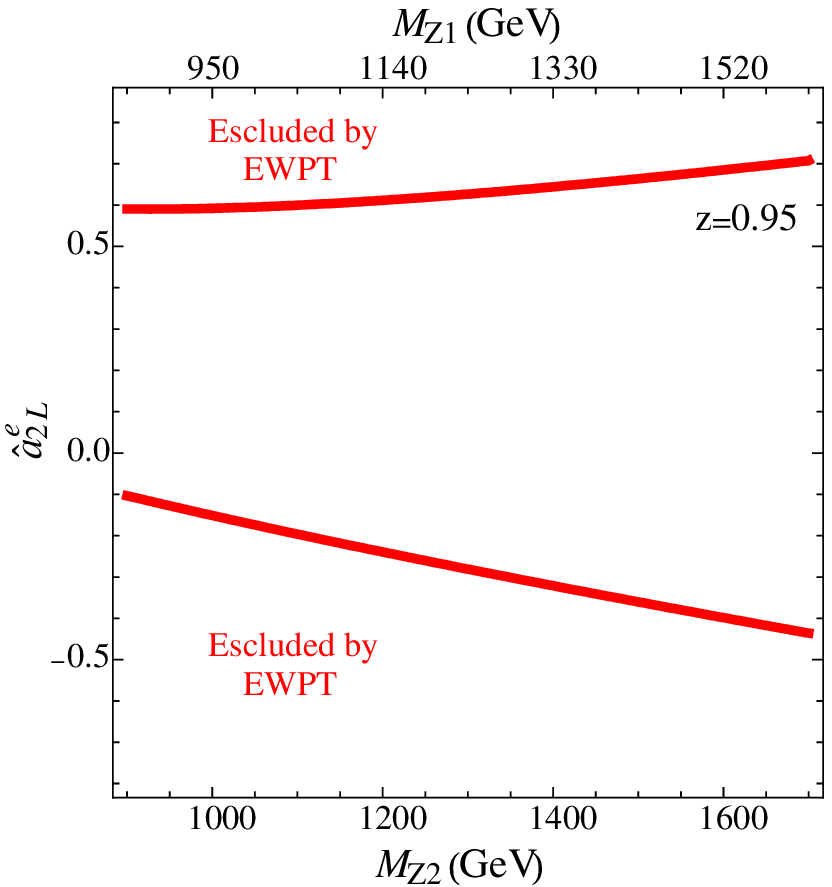,width=5.5cm}}
\end{picture}
\end{center}
\vskip 6.cm
\caption{95\% C.L. EWPT bounds in the plane ($M_{Z_2}, \hat{a}_{2L}^\Pe$). The 
allowed region is inside the thick red curve; lower and upper values for 
$M_{Z_2}$ come respectively from approximation and unitarity limits (see 
Fig.~\ref{bi1}). The four plots refer to four different values of the free 
$z$-parameter: $z$=0.4 (top-left), $z$=0.6 (top-right), $z$=0.8 
(bottom-left), $z$=0.95 (bottom-right).}
\label{fig:m2a2l}
\end{figure}

The real EWPT constraint comes from the $\epsilon_3$ parameter, which imposes 
a strict relation between $\hat{a}^\Pe_{1L}$ and $\hat{a}^\Pe_{2L}$, as 
displayed by the dashed-line contour in Fig.~\ref{bi1} (right panel) where
the band width is due to the experimental error on $\epsilon_3$.
This strong bound allows one to derive a second relation between the free 
parameters of the model. Assuming $\epsilon_3$ equal to its central value
\cite{Barbieri:2004qk}:
\be\label{eps3}
\epsilon_3=\frac{\sqrt{2}e}{g_1} ({\hat{a}_{1L}^e-z^2
\hat{a}_{2L}^e})-\frac{e^2}{g_1^2 c^2_\theta} (1+z^4)=
4.8 \times 10^{-3}
\ee 
one can write the $Z_1$-boson electron coupling, $\hat{a}_{1L}^\Pe$, as a 
function of $\hat{a}_{2L}^\Pe$ at fixed $Z_{1,2}$-boson masses. The net 
result is
that the number of independent free parameters gets further reduced to three. 
In the next sections, to describe the four-site Higgsless model, we will 
choose the following set: $z$, $M_{Z_2}$, $\hat{a}_{2L}^\Pe$ (from 
Eqs.(\ref{A14}),(\ref{eq:M2}),(\ref{A16}) 
$z=M_1/M_2=M_{Z_1}/M_{Z_2}+\mathcal{O}(x^2)$).
The parameter space allowed by EWPT is shown in Fig.~\ref{fig:m2a2l} 
for various values of the free $z$ parameter.

%%%%%%%%%%%%%%%%%%%%%%%%%%%%%%%%%%%%%%%%%%%%%%%%%%%%%%%%%%%%%%%%%%%%%%%%%%%%%%%%%%%%%%%%%%%%%%%%%%%%%%%%%%%%%%%%%%%%%

\section{Extra $Z_{1,2}$-bosons: Mass Spectrum, Decay Widths and Branching 
Ratios}
\label{spectrum}
In this section, we summarize the main properties of the heavy 
$Z_{1,2}$-bosons. The first peculiarity of the four-site model is related to 
the nature of the two extra gauge bosons and their mass hierarchy. The 
lighter particle, $Z_1$, is a vector boson while the heavier one, $Z_2$, is an  axial-vector (neglecting electroweak corrections). Oppositely to closely related models, like the walking 
technicolor  \cite{Belyaev:2008yj}, no mass spectrum inversion is possible. 
The mass splitting, $\Delta M=M_{Z_2}-M_{Z_1}$, is always positive and its size 
depends on the free $z$-parameter: 
\be
\Delta M=(1-z)M_{Z_2}+\mathcal{O}(x^2)\qquad 0<z<1.
\ee  
We can thus have scenarios where the two resonances lie quite apart from each 
other, and portions of the parameter space in which they are (almost) 
degenerate. 
In the latter case, the multi-resonance distinctive signature would collapse 
into the more general single $Z^\prime$ signal. The four-site model would thus 
manifest a degeneracy with the well known extra U(1) theories predicting only 
one additional neutral gauge boson.
The mass spectrum has both a lower and an upper bound, as discussed in 
Sect.\ref{EWPT_UNIT}. It lies roughly in the range 350$\le$ M$\le$ 3000 GeV.

\begin{figure}[!htbp]
\begin{center}
\unitlength1.0cm
\begin{picture}(7,6)
\put(-3.7,1.8){\epsfig{file=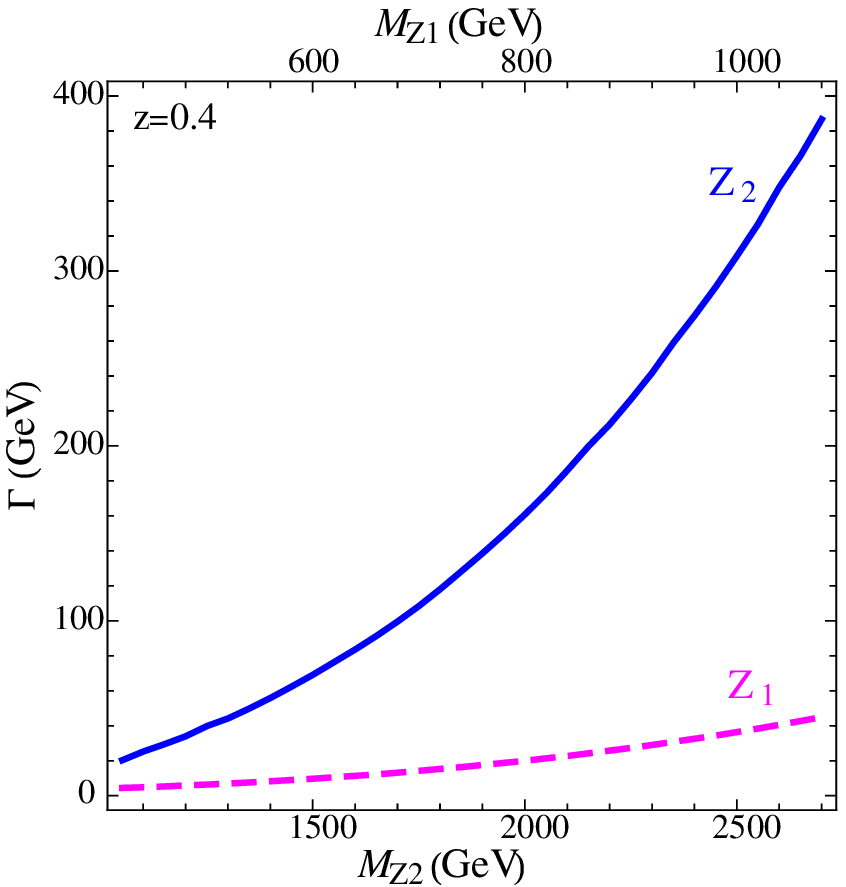,width=5.5cm}}
\put(3.9,1.8){\epsfig{file=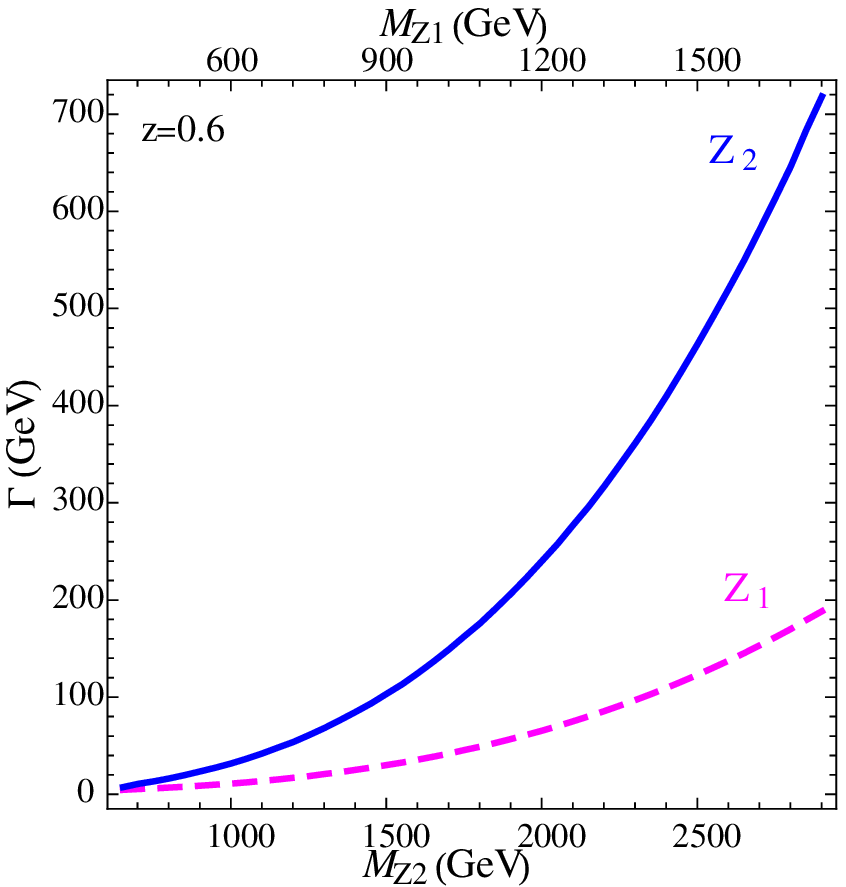,width=5.5cm}}
\put(-3.7,-4.3){\epsfig{file=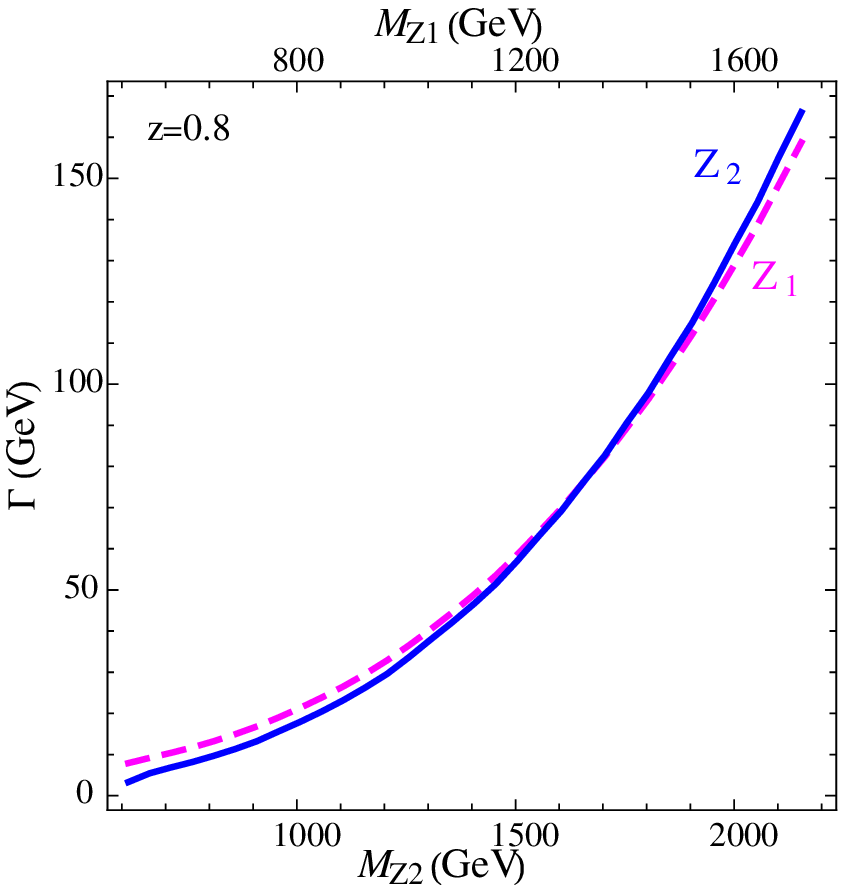,width=5.5cm}}
\put(3.9,-4.3){\epsfig{file=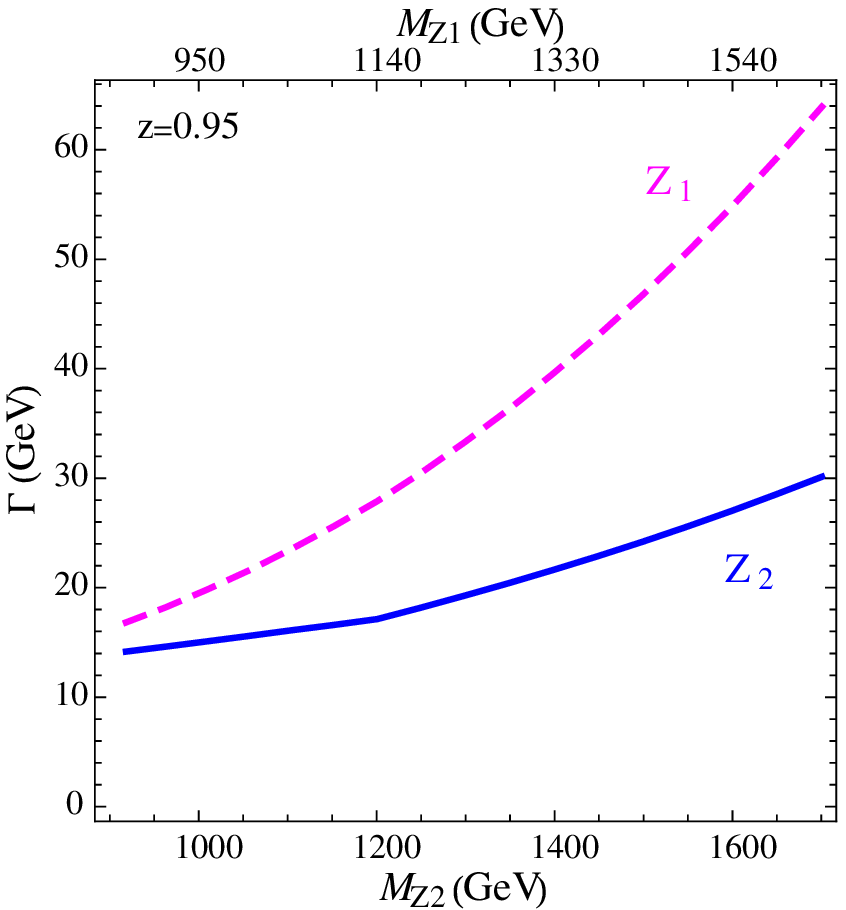,width=5.5cm}}
\end{picture}
\end{center}
\vskip 4.cm
\caption{Total decay width of $Z_1$ (magenta dashed line) and $Z_2$ (blue 
solid line) extra gauge bosons as a function of their mass, $M_{Z_1}$ 
(upper x-axis) and $M_{Z_2}$ (lower x-axis). The four plots refer to four
different values of the free $z$-parameter: $z$=0.4 (top-left), $z$=0.6 (top-right),
 $z$=0.8 (bottom-left), $z$=0.95 (bottom-right).}
\label{fig_totalwidth}
\end{figure}

The total widths of the two heavy gauge bosons, $\Gamma_{Z_{1,2}}$, are 
displayed in Fig.~\ref{fig_totalwidth} as a function of their mass for four 
values of the $z$-parameter: $z=0.4, 0.6, 0.8, 0.95$. They have been computed 
by taking as $Z_{1,2}f\bar f$ couplings those corresponding to the maximal
$\hat a^\Pe_{2L}$ value allowed by EWPT for each $z$ value (upper lines in
Fig. \ref{fig:m2a2l}). This maximizes the fermionic contribution to 
the total decay width, and it will be used later to show the maximal 
branching ratio one might expect for the $Z_{1,2}$-boson decay into electrons.

From Fig.~\ref{fig_totalwidth}, one can see that both $Z_1$ and $Z_2$ are 
very narrow for low mass values. In the low edge of their spectrum, the 
magnitude of their total width is around a few GeV. Since the width is 
dominated by the $Z_{1,2}$-boson decay into gauge boson pairs (when 
kinematically allowed) whose behaviour is proportional to $M_{Z_1,Z_2}^3$,
it then increases with the mass up to hundreds of GeV. This can be easily 
seen from the diboson contribution to the total decay widths which, at leading 
order in the $x$-parameter given in Eq.~(\ref{x-parameter}), have the following
expressions
\be
\Gamma_{Z_1}^{WW}=\frac{1}{3\pi}\left (\frac{\gt}{16}\right )^2\frac{M_{Z_1}^3}
{M_W^2}(1-z^4)(1+z^2)
\ee
\be
\Gamma_{Z_2}^{W_1W}=\frac{1}{3\pi}\left (\frac{\gt}{16}\right )^2\frac{M_{Z_2}^3}
{M_W^2} z^4 (1-z^2)^3 \left[1+10z^2+z^4\right].
\ee
for the $Z_1$ and $Z_2$ bosons, respectively.

Let us briefly comment these formulas. The dependence of $\Gamma_{Z_{1,2}}$ 
from the $z$-parameter has a twofold source.
It comes partially from the structure of the $Z_1WW$ and $Z_2W_1W$ trilinear 
couplings (see Appendix \ref{appendixA}), and in part by the fact that the $z$-parameter 
determines whether the $Z_2\rightarrow W_1W$ channel is kinematically open 
(the $Z_2\rightarrow W_1W_1$ channel is suppressed by the coupling).
At fixed mass, $\Gamma_{Z_1}^{WW}$ slightly decreases by increasing the 
$z$-parameter owing to the self-coupling $a_{WWZ_1}\propto (1-z^4)$ given in eq. (\ref{AA4}). On the 
contrary, $\Gamma_{Z_2}^{W_1W}$ gets 
larger by increasing the $z$-parameter owing to the trilinear coupling 
$a_{W_1WZ_2}\propto z^2$ given in eq. (\ref{AA7}). This latter behaviour persists until the growth of 
the coupling can balance the reduction of the kinematically allowed phase 
space for producing the $W_1W$ boson pairs. As the $z$-parameter increases, 
$Z_2$ and 
$W_1$ become more and more degenerate in mass: $M_{Z_2}\simeq M_{W_1}$. 
Hence, the phase space shrinks until it gets closed for $z\ge 1-M_W/M_{Z_2}$. 
When the $Z_2\rightarrow W_1W$ channel is kinematically forbidden, the total 
width is given by the $Z_2$-boson decay into fermions. Its size is thus 
drastically reduced as displayed in Fig.~\ref{fig_totalwidth} (bottom-right) 
for $z=0.95$. 
The above mentioned combination of coupling and phase space effects explains 
the $\Gamma_{Z_{1,2}}$ behaviour shown in Fig.~\ref{fig_totalwidth}. From the 
plots, it is also 
clear that $Z_2$ is broader than $Z_1$ as soon as its decay into boson pairs 
opens up. 

\begin{figure}[!htbp]
\begin{center}
\unitlength1.0cm
\begin{picture}(8,7)
\put(-3.3,0.6){\epsfig{file=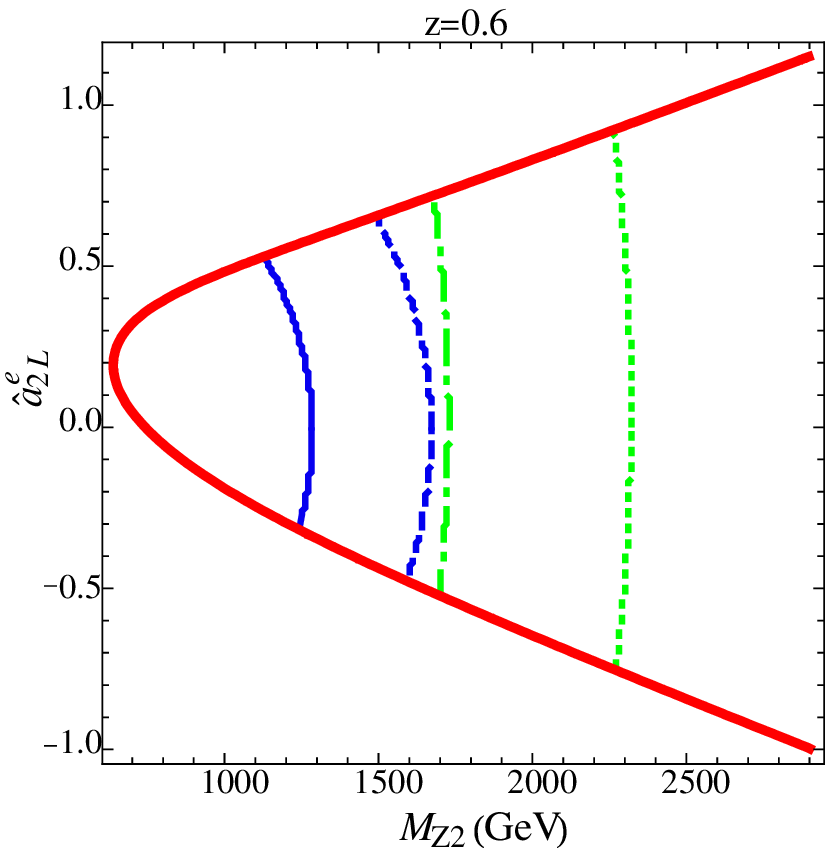,width=6.6cm}}
\put(4.0,0.6){\epsfig{file=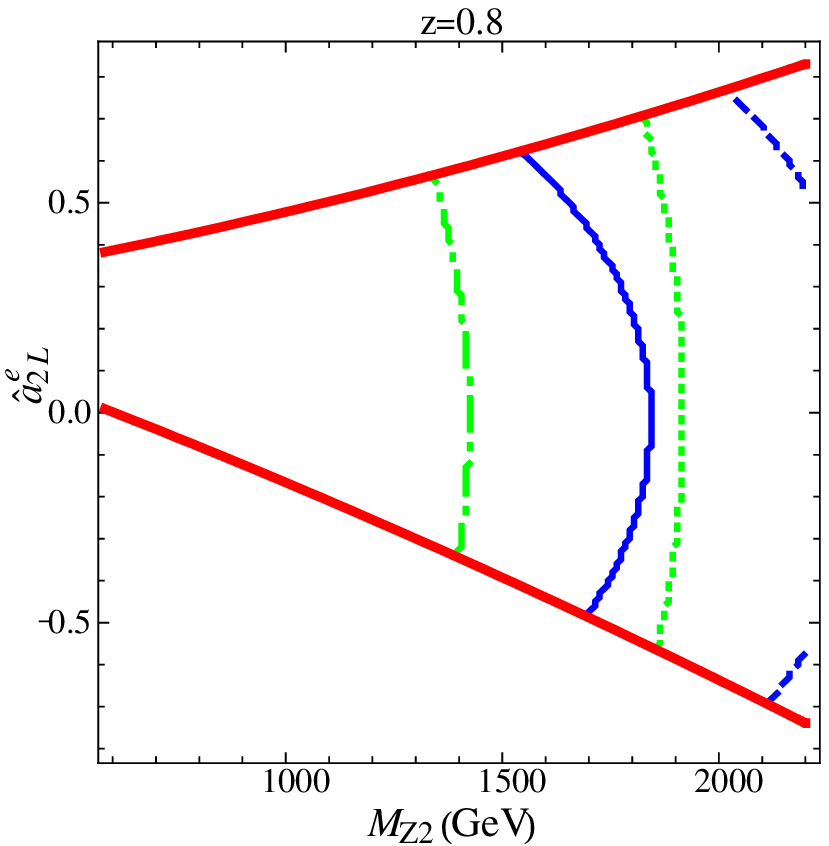,width=6.6cm}}
\end{picture}
\end{center}
\vskip -1.cm
\caption{Left: Contour plots in the plane $(M_{Z_2}$, $\hat a_{2L}^\Pe)$ at 
fixed $z=0.6$ representing the following requirements: 
$\Gamma_{Z_2}/2= R_{LHC}$ (blue solid), $\Gamma_{Z_1}/2= R_{LHC}$ (green 
dot-dashed), $\Gamma_{Z_2}/2= R_{TEV}$ (blue dashed), and 
$\Gamma_{Z_1}/2= R_{TEV}$ (green dotted) as explained in the text. 
Right: same as left-panel for $z=0.8$.
The total width can be measured in the righthand regions of the corresponding 
curve.}
\label{fig_widthcontour}
\end{figure}

Since the $Z_{1,2}$ total width can range between a few and hundreds GeV, a 
natural question is whether it would be possible to measure it at the LHC.
The mass resolution during the early stage of the LHC is estimated to be 
$R_{LHC}=2\%$M \cite{Basso:2010pe,CMSdet,ATLASdet}. If 
$0.5\cdot\Gamma_{Z_{1,2}}\ge R_{LHC}$ (or $\Gamma_{Z_{1,2}}/(2M_{Z_1,Z_2})\ge 
2\%$), then the shape of the corresponding resonance could be fully 
reconstructed and analysed. In this case, the decay width could be measured. 
In Fig.~\ref{fig_widthcontour}, we show  contour plots defining in what 
region of the parameter space the total width could be in principle measured. 
The two plots refer to two representative values of the free $z$-parameter: 
$z=0.6, 0.8$. Within the corresponding parameter space, limited by EWPT 
bounds (thick red lines), the figures contain four contour lines. The blue 
solid line corresponds to $0.5\cdot\Gamma_{Z_2}= R_{LHC}$, while 
the green dot-dashed one refers to $0.5\cdot\Gamma_{Z_1}= R_{LHC}$. In the 
righthand region of each curve, the corresponding total width could be 
measured.  
For comparison, we show also the same contour plots for the Tevatron assuming 
a mass resolution $R_{TEV}=3.4\%$M \cite{Basso:2010pe}. These are represented 
by the blue dashed and green dotted lines for the $Z_2$ and $Z_1$ bosons, 
respectively.  
As one can see, the $\Gamma_{Z_1}$-contour is almost independent on the 
$Z_1$-boson coupling to fermions. The contour line simply shifts (very 
slightly) to higher mass values by increasing the $z$-parameter just because 
$\Gamma_{Z_1}$ decreases correspondingly at fixed mass. The behaviour of the 
$\Gamma_{Z_2}$-contour is 
instead more complicated. It in fact depends on whether or not the decay into 
boson pairs is allowed. More the $Z_2\rightarrow W_1W$ phase space shrinks, 
more the decay into fermions becomes important. The contour lines acquire 
therefore a dependence on the $Z_2$-boson coupling to ordinary matter, 
represented in the plot by the $Z_2$-boson coupling to left-handed electrons 
$\hat a_{2L}^\Pe$, which gets enhanced with the $z$-parameter. Moreover, by 
increasing the $z$-parameter, the contour lines shift sensibly to higher mass 
values, as for the $Z_1$-boson case. This trend continues until the 
$Z_{1,2}$-boson widths get
smaller than the mass resolution at both LHC and Tevatron in the full 
parameter space (i.e. all contour lines shift beyond the mass range allowed 
by the perturbative unitarity constraint). This is represented by the 
$z=0.95$ scenario, which we do not explicitly display.    

The possibility of measuring the width of the extra gauge bosons, at least in 
a sizeable part of the parameter space, is a distinctive feature of the 
four-site model. Common extra U(1) theories 
present in the literature predict indeed an additional $Z^\prime$-boson purely 
decaying into fermion pairs. As a consequence, the $Z^\prime$ width is 
generally expected to be very narrow and below the foreseen mass resolution
at the LHC.
  
\begin{figure}[!htbp]
\begin{center}
\unitlength1.0cm
\begin{picture}(7,6)
\put(-4.1,1.3){\epsfig{file=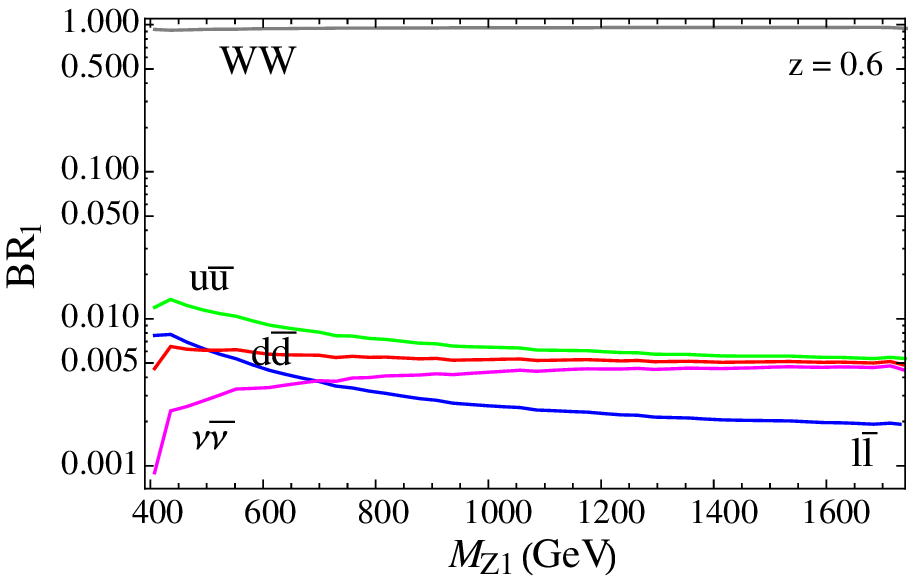,width=7.5cm}}
\put(3.5,1.3){\epsfig{file=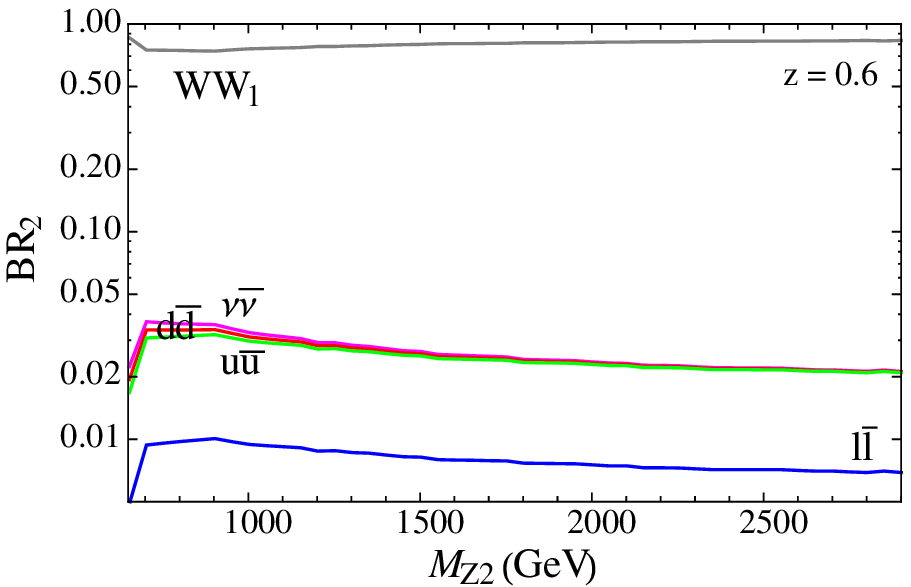,width=7.5cm}}
\put(-4.1,-3.7){\epsfig{file=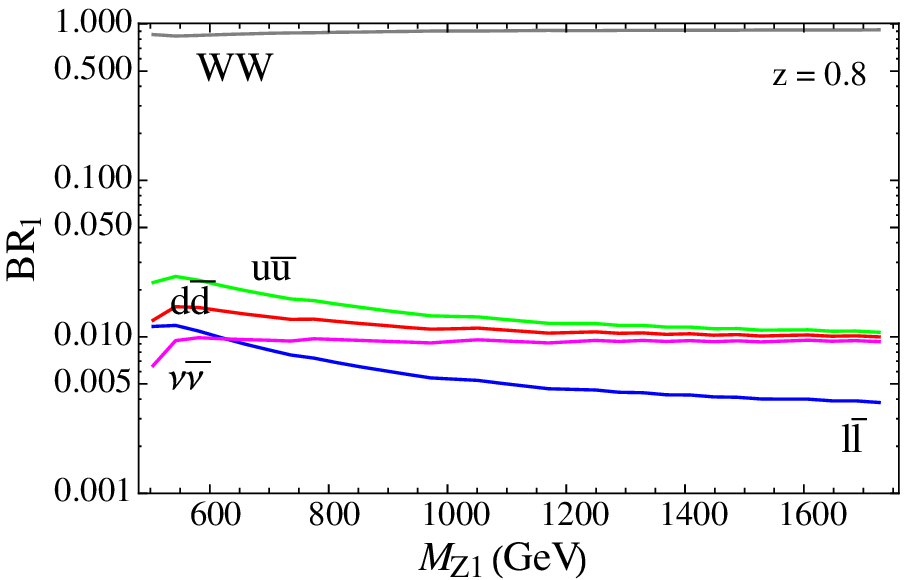,width=7.5cm}}
\put(3.5,-3.7){\epsfig{file=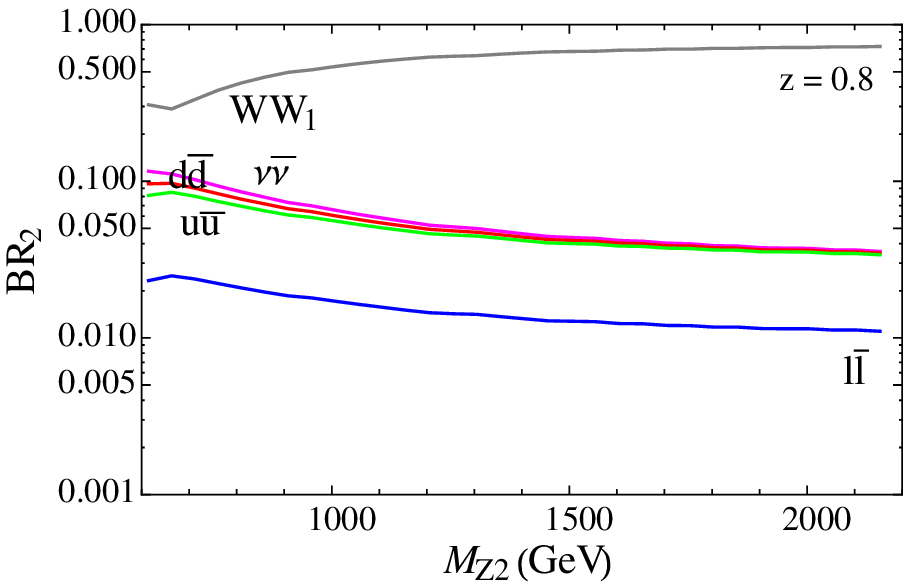,width=7.5cm}}
\put(-4.1,-8.7){\epsfig{file=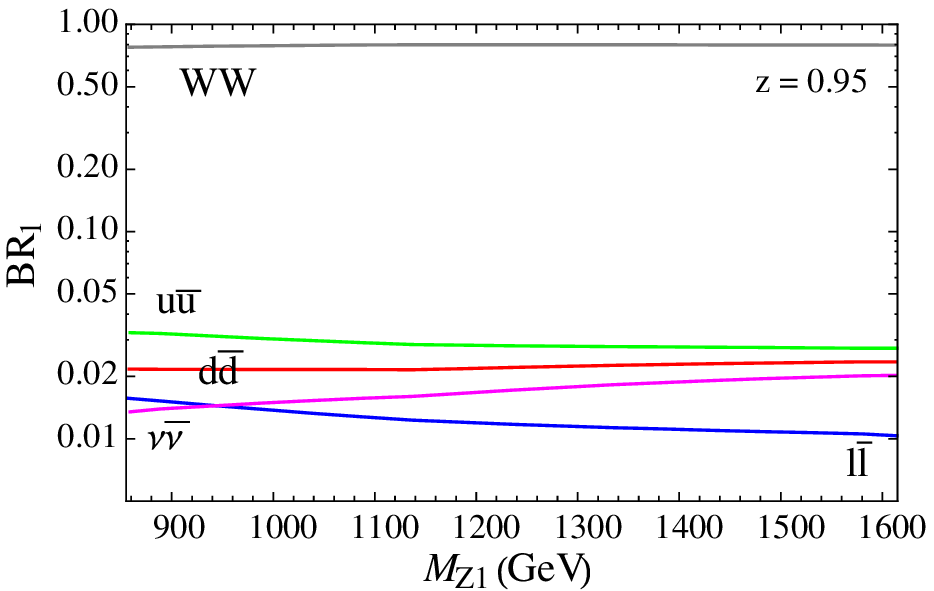,width=7.5cm}}
\put(3.5,-8.7){\epsfig{file=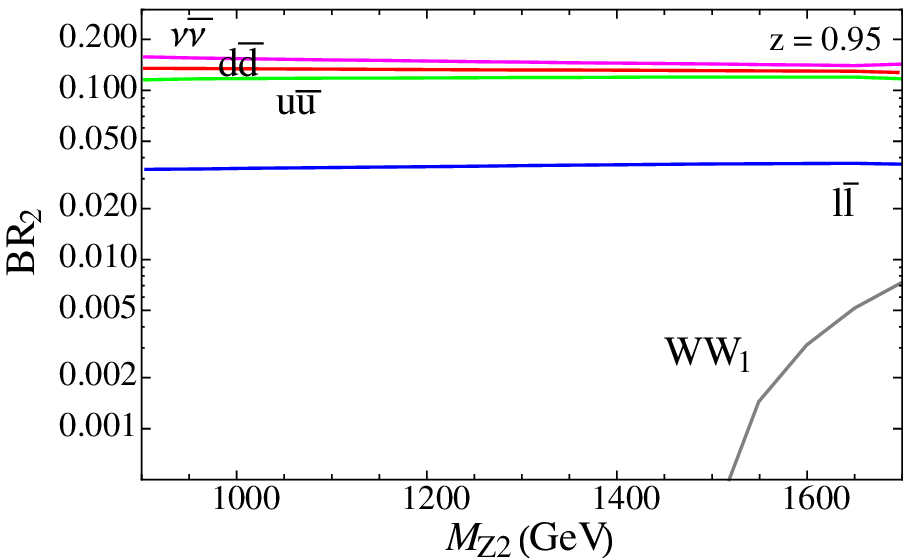,width=7.5cm}}
%\put(-4.1,-13.3){\epsfig{file=BR1_z95.eps,width=7.5cm}}
%\put(3.5,-13.3){\epsfig{file=BR2_z95.eps,width=7.5cm}}
\end{picture}
\end{center}
\vskip 8.cm
\caption{Left: $Z_1$-boson branching ratios as a function of its mass. 
Right: $Z_2$-boson branching ratios versus its mass. From top to bottom, the 
free $z$-parameter assumes the three representative values: 
$z=0.6, 0.8, 0.95$.}
\label{fig_brs}
\end{figure}

The $Z_{1,2}$ branching ratios are shown in Fig.~\ref{fig_brs} for three 
values of the $z$-parameter: $z$=0.6,0.8,0.95. Here we do not show the $z$=0.4 case as it does not differ 
sensibly from the $z$=0.6 scenario. The $Z_{1,2}$-boson BRs into 
fermion pairs are evaluated by considering, for each mass value, the 
maximum coupling allowed by EWPT bounds. Let us discuss first the $Z_1$ 
branching ratios displayed
on the leftside plots. The lighter extra gauge boson can decay into 
fermions and dibosons: $Z_1\rightarrow f\bar f$ and $Z_1\rightarrow WW$. 
In the figure, 
the branching in down-quarks, up-quarks, neutrinos (summed up over all 
generations), electrons and W-boson pairs is shown. For all $z$-values, the 
diboson channel is the dominant one. The decay into electrons is below $2\%$
(slightly increasing with the $z$-parameter), but it is anyhow competing if 
one relies on clean purely leptonic final states as 
$BR(Z_1\rightarrow WW\rightarrow ee\nu_e\nu_e )=BR(Z_1\rightarrow WW)/81$.

The $Z_2$ branching ratios are displayed on the right. The heavier extra 
gauge boson is purely axial up to electrowek corrections, hence it can decay into fermions and mixed 
diboson pairs: $Z_2\rightarrow f\bar f$ and $Z_2\rightarrow W_1W$ (
$Z_2\rightarrow WW$ and $Z_2\rightarrow W_1W_1$ are highly suppressed).
Once more, the competing coupling and phase space effects play a delicate role.
Even if the decay into electron-pairs is always below $4\%$, the global 
branching ratio into fermions can become dominant. By increasing the 
$z$-parameter, the phase space for the mixed diboson decay gets reduced or 
even disappears, and the fermions take over as shown clearly in the 
bottom-right plot ($z$=0.95).   

\section{Drell-Yan production at the LHC and the Tevatron}\label{dy} 

We can now consider the production of the two neutral gauge bosons, 
$Z_{1,2}$, predicted by the four-site Higgsless model at the LHC and the 
Tevatron through the Drell-Yan channel. 
Owing to the introduction of direct couplings between 
ordinary matter and extra gauge bosons, in addition to the usual indirect 
ones due to the mixing, the experimental bounds from electroweak precision 
data on the model parameters are indeed less stringent. As a consequence,
and in contrast with the existing fermiophobic Higgsless literature,
quite large couplings between SM fermions and extra gauge bosons are
allowed (see Fig.~\ref{fig:m2a2l}).

\subsection{Processes and their computation}
\label{se:processes}

We analyze in detail the neutral Drell-Yan channels
\be
\Pp\Pp\to \Pe^+\Pe^-,~~~~~~~ \Pp\bar\Pp\to \Pe^+\Pe^-
\ee
at the LHC and the Tevatron, respectively. The two channels differ only by the 
initial state, and are characterized by an isolated electron and positron in 
the final state.
These processes can involve the production of the two additional gauge bosons, 
$Z_1$ and $Z_2$, as intermediate states. They are described by the generic 
formula
\beqar \rd\si^{h_1 h_2}(P_1,P_2,p_f) = \sum_{i,j}\int\rd x_1 \rd
x_2~ f_{i,h_1}(x_1,Q^2)f_{j,h_2}(x_2,Q^2)
\,\rd\hat\si^{ij}(x_1P_1,x_2P_2,p_f), \eeqar where $p_f$ summarizes
the final-state momenta, $f_{i,h_1}$ and $f_{j,h_2}$ are the
distribution functions of the partons $i$ and $j$ in the incoming
hadrons $h_1$ and $h_2$ with momenta $P_1$ and $P_2$, respectively,
$Q$ is the factorization scale, and $\hat\si^{ij}$ represent the
cross sections for the partonic processes. At the LHC, since the two incoming
hadrons are protons and we sum over final states with opposite
charges, we find
\beqar\label{eq:convol_neut_LHC}%\refeq{eq:convol}
\rd\si^{h_1h_2}(P_1,P_2,p_f) = \int\rd x_1 \rd x_2 &&\sum_{Q=u,c,d,s,b}
\Bigl[f_{\bar\PQ,\Pp}(x_1,Q^2)f_{\PQ,\Pp}(x_2,Q^2)\,\rd\hat\si^{\bar\PQ\PQ}
(x_1P_1,x_2P_2,p_f)
\nl&&{}
+f_{\PQ,\Pp}(x_1,Q^2)f_{\bar\PQ,\Pp}(x_2,Q^2)\,\rd\hat\si^{\PQ\bar\PQ}
(x_1P_1,x_2P_2,p_f)\Bigr].
\eeqar
At the Tevatron, since the two incoming hadrons are proton and anti-proton, 
the same observable reads instead as
\beqar\label{eq:convol_neut_TEV}%\refeq{eq:convol}
\rd\si^{h_1h_2}(P_1,P_2,p_f) = \int\rd x_1 \rd x_2 &&\sum_{Q=u,c,d,s,b}
\Bigl[f_{\bar\PQ,\Pp}(x_1,Q^2)f_{\PQ,\bar\Pp}(x_2,Q^2)\,\rd\hat\si^{\bar\PQ\PQ}
(x_1P_1,x_2P_2,p_f)
\nl&&{}
+f_{\PQ,\Pp}(x_1,Q^2)f_{\bar\PQ,\bar\Pp}(x_2,Q^2)\,\rd\hat\si^{\PQ\bar\PQ}
(x_1P_1,x_2P_2,p_f)\Bigr].
\eeqar

The tree-level amplitudes for the partonic processes have been generated by
means of {\tt PHACT} \cite{Ballestrero:1999md}, a set of routines based on the
helicity-amplitude formalism of \citere{Ballestrero:1994jn}. The matrix
elements have been inserted in the Monte Carlo event generator
{\tt FAST$\_$2f}, dedicated to Drell-Yan processes at the EW and QCD leading
order. {\tt FAST$\_$2f} can compute simultaneously the new-physics signal and
the SM background. It can generate cross-sections and distributions for
any observable, including any kind of kinematical cuts. The code is
moreover interfaced with {\tt PYTHIA} \cite{Sjostrand:2006za}. This feature
can allow a more realistic analysis, once {\tt FAST$\_$2f} is matched
with detector simulation programs. 

\subsection{Numerical setup}
\label{se:setup}

For the numerical results presented here, we have used the following
input values \cite{Yao:2006px}: $M_Z=91.187\GeV$, $\GZ=2.512\GeV$, $\GW
=2.105\GeV$,  $\alpha(M_Z)=1/128.88$, $G_F = 1.166\times 10^{-5}$ GeV$^{-2}$.
In our scheme, the weak mixing-angle and the $W$-boson mass are derived 
quantities. We use  the fixed-width scheme for the matrix element evaluation, and   the CTEQ6L\cite{Pumplin:2002vw} for the parton 
distribution functions at the factorization scale:
\begin{equation}
Q^2=M_{\inv}^2(\Pe^+\Pe^-)
\end{equation}
where $M_{\inv}$ denotes the invariant mass. This scale choice appears to be 
appropriate for the calculation of differential cross sections, in particular 
for lepton distributions at high energy scales.

When considering the DY-channel at the LHC, we have moreover implemented a 
general set of acceptance cuts appropriate for LHC analyses, and defined as 
follows:
\begin{itemize}
\item {lepton transverse momentum $\PT(l)>20\GeV$},

\item {lepton pseudo-rapidity $|\eta_l |< 2.5$},
where $\eta_l=-\log\left (\tan\theta_l/2\right )$, and
  $\theta_l$ is the polar angle of the charged lepton $l$
with respect to the beam.
\end{itemize}
For the process at hand, we have also used further cuts which are described 
in due time. We assume as electron detection efficiency 
$\epsilon_\Pe =90\%$ \cite{CMSdet} and we present
results for the 7 TeV LHC with an integrated luminosity L$=1\fba^{-1}$ and for
 the 14 TeV LHC with an integrated luminosity L$=10\fba^{-1}$.  

For the study of the Drell-Yan channel at the Tevatron, we assume instead an 
overall signal acceptance of the order of $20\%$ \cite{D0preliminary}. 
Moreover, we consider two values for the integrated luminosity: 
$L=3.6\fba^{-1}$ (recent preliminary D0 analysis) and $L=10\fba^{-1}$ (next 
two year projection).

\subsection{$Z_{1,2}$-boson production at the LHC and the Tevatron}
\label{se:boson_production}

The additional gauge bosons, predicted by the four-site Higgsless model, can 
be produced at the LHC via the DY process 
$pp\rightarrow\gamma, Z, Z_{1,2}\rightarrow\Pe^+\Pe^-$. 

\begin{figure}[!htbp]
\begin{center}
\unitlength1.0cm
\begin{picture}(7,6)
\put(-4.1,1.8){\epsfig{file=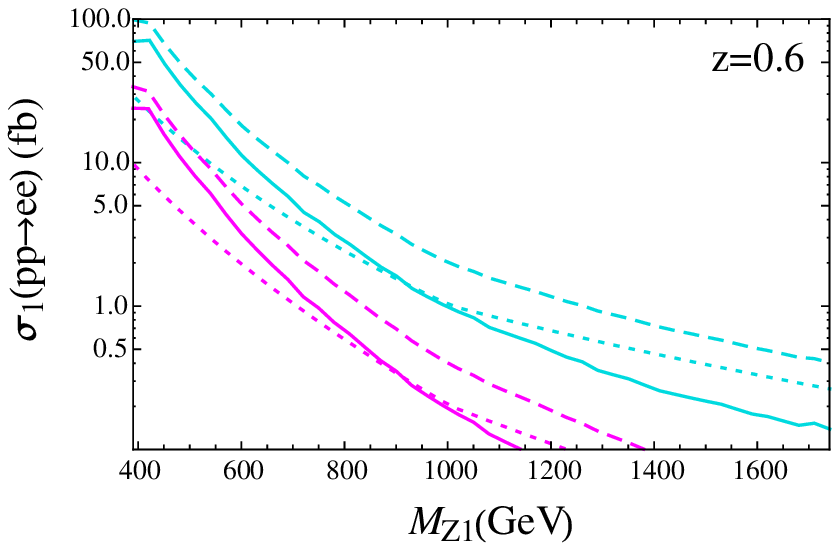,width=7.5cm}}
\put(3.5,1.8){\epsfig{file=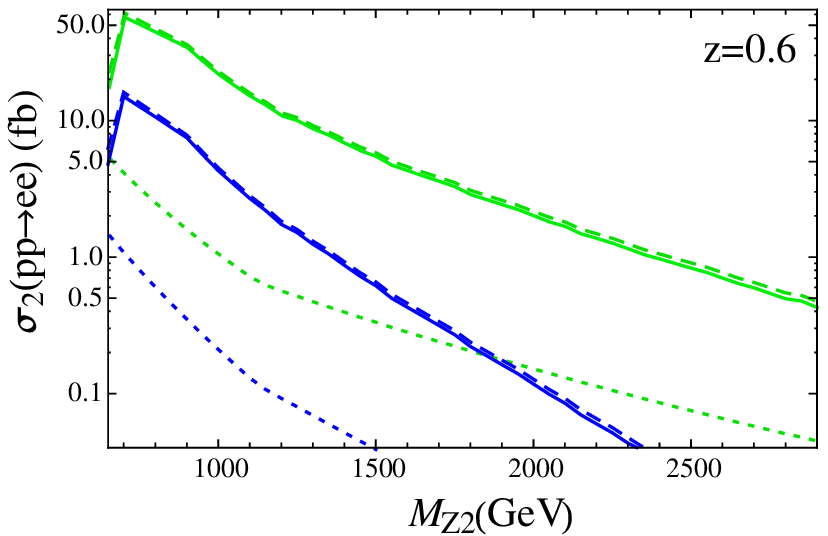,width=7.5cm}}
\put(-4.1,-3.6){\epsfig{file=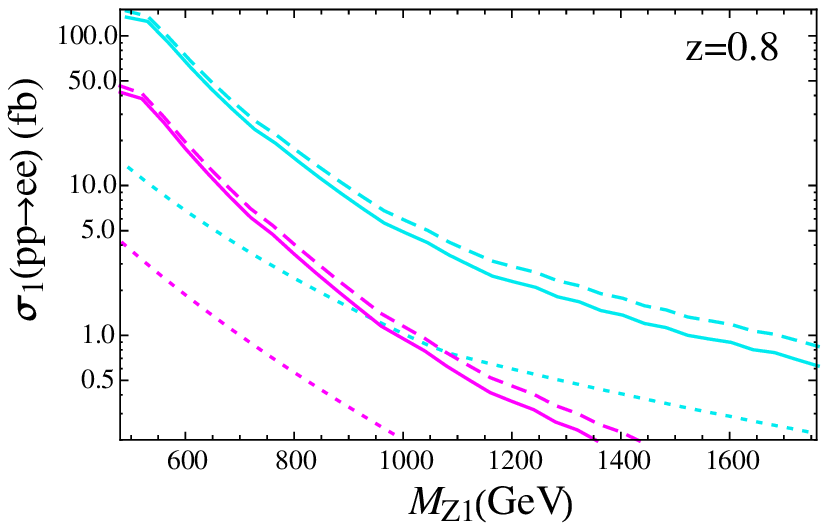,width=7.5cm}}
\put(3.5,-3.6){\epsfig{file=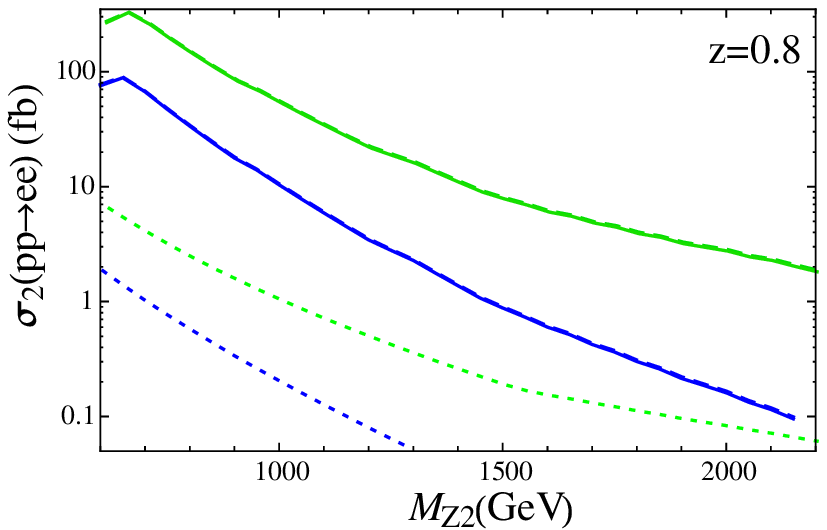,width=7.5cm}}
\put(-4.1,-8.6){\epsfig{file=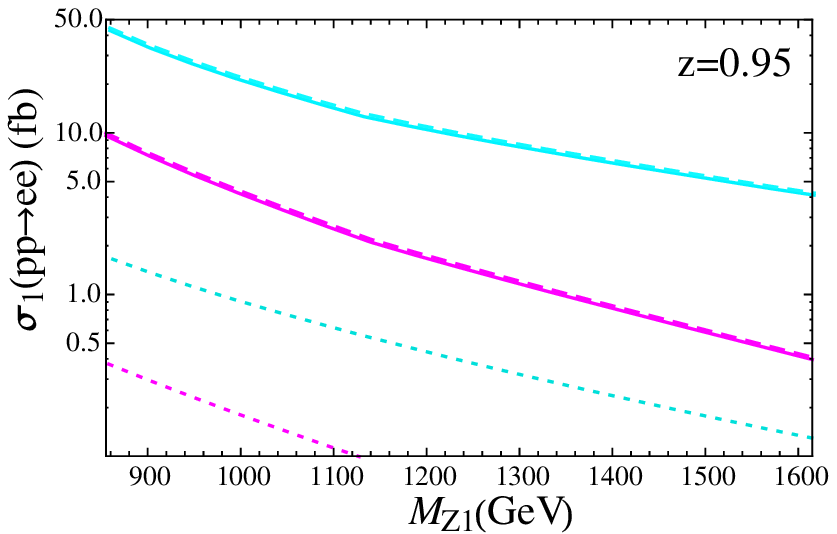,width=7.5cm}}
\put(3.5,-8.7){\epsfig{file=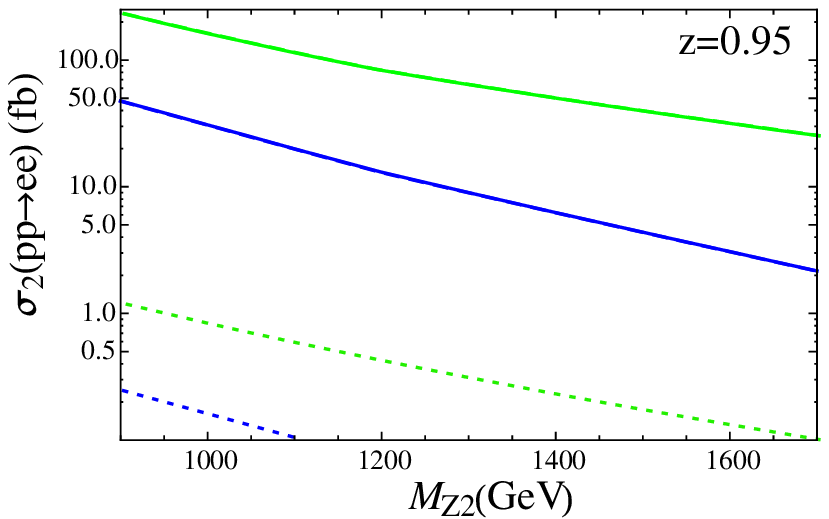,width=7.4cm}}
\end{picture}
\end{center}
\vskip 8.cm
\caption{Left: $Z_1$-boson cross-sections for the maximal $\hat a^\Pe_{2L}$ 
allowed by EWPT, in the mass window given in Eq.(\ref{window}), as a function 
of $M_{Z_1}$ at the 7 TeV LHC (lower curves) and the 14 TeV LHC (upper 
curves). The dashed lines represent the total cross-sections (T), including 
the interference of the $Z_1$-boson signal with both the SM background and the 
$Z_2$-boson signal. The dotted lines show the SM backgrounds (B). The solid 
lines correspond to their differences: S=T-B.  
Right: same for the $Z_2$-boson. From top to bottom, the free $z$-parameter 
assumes the three representative values: $z$=0.6, 0.8, 0.95.}
\label{fig_cross-section_LHC}
\end{figure}

The corresponding 
total cross-sections, in the gauge boson mass interval allowed by 
unitarity bounds, are shown in Fig.~\ref{fig_cross-section_LHC} for three 
representative values of the free $z$-parameter: $z$=0.6, 0.8, 0.95. These are 
just bare values, useful only to give an idea of 
the magnitude of the expected cross-sections under the resonances.
The displayed cross-sections have been in fact computed within the symmetric 
mass window, 
\be
|M_{\inv}(\Pe^+\Pe^-)-M_{Z_1,Z_2}|\le max(\Gamma_{Z_1,Z_2}/2, R)
\label{window}
\ee 
with $R$ the extimated mass resolution for LHC and Tevatron. 
%, and with no acceptance cuts. Also the electron detection efficiency is not included. 
Moreover, they have been calculated for the maximum value of the 
$Z_2$-boson coupling to left-handed electrons allowed by the EWPT for the given 
$z$-parameter. 

The left-side panel of Fig.~\ref{fig_cross-section_LHC} shows the total 
cross-section under the $Z_1$-resonance. The right-side panel refers to the
$Z_2$-boson. 
For each $z$ value we plot the cross section, corresponding to the maximum 
coupling $\hat a^\Pe_{2L}$ allowed by EWPT, for $\sqrt{s}=$ 7 and 14 TeV.
At fixed mass, $M_{Z_1,Z_2}$, the $Z_{1,2}$-boson cross-sections 
get larger by increasing the value of the $z$-parameter. This 
effect is due to the fact that high $z$-values help in relaxing EWPT 
constraints on the $Z_{1,2}$-boson coupling to SM fermions, as can be seen in 
Fig.~\ref{fig:m2a2l} which shows the EWPT bounds on the 
representative $Z_2$-boson coupling to electrons, ${\hat a}^\Pe_{2L}$.

\begin{figure}[!htbp]
\begin{center}
\unitlength1.0cm
\begin{picture}(7,6)
\put(-4.5,-1.5){\epsfig{file=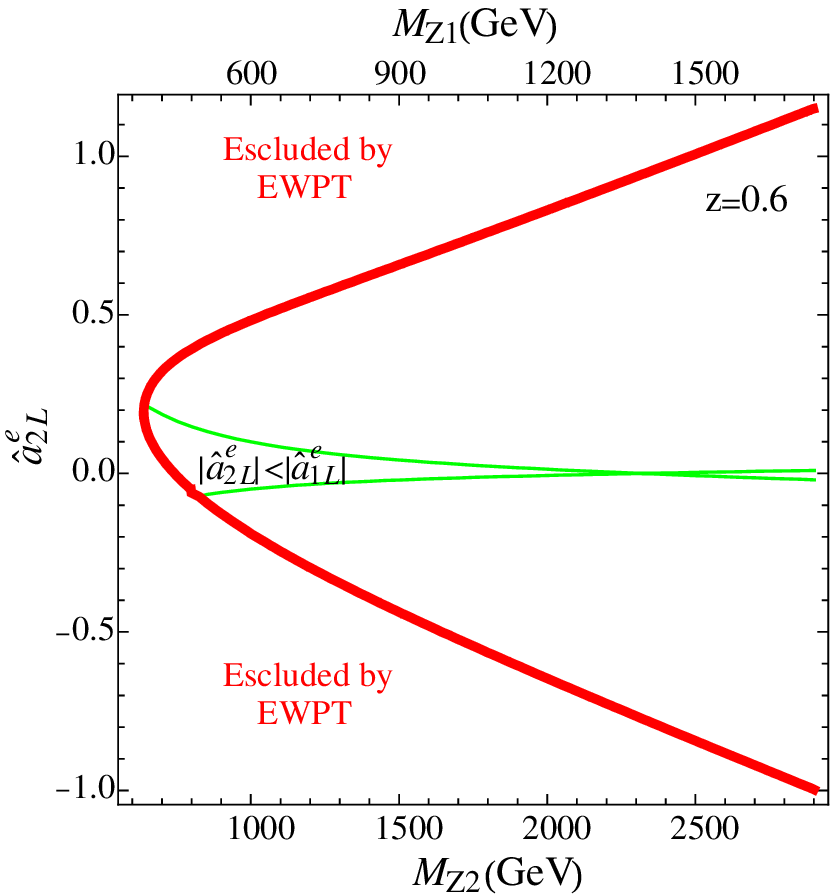,width=7.cm}}
\put(3.5,-1.5){\epsfig{file=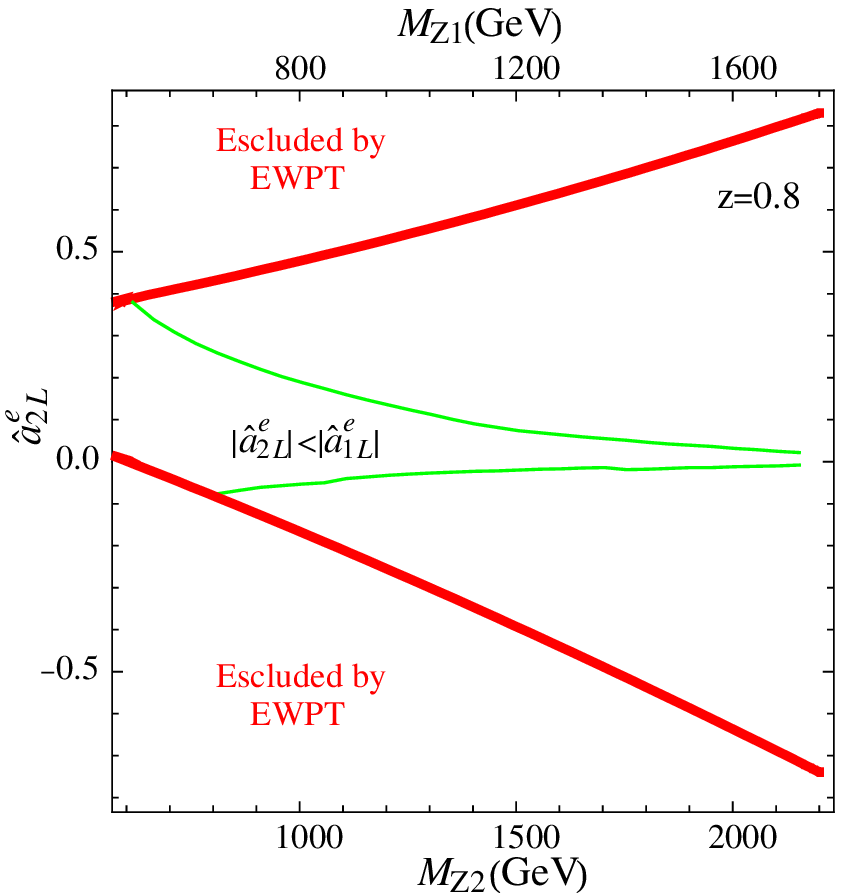,width=7.cm}}
\end{picture}
\end{center}
\vskip 1.cm
\caption{The red thick solid line gives the parameter space allowed by EWPT. 
The green solid lines delimit the region where the $Z_2$-boson coupling to 
electrons (${\hat a}^\Pe_{2L}$) is smaller than the $Z_1$-boson coupling to 
electrons (${\hat a}^\Pe_{1L}$). We consider two sample values of the free 
$z$-parameter: $z$=0.6 (left-panel) and $z$=0.8 (right-panel).}
\label{fig:a1_a2}
\end{figure}

Moreover, while the total cross-section under the $Z_1$-resonance is at most 
$\sigma_1\simeq$ 50 fb at the 7 TeV LHC, the $Z_2$-boson cross-section can be 
sensibly higher, of the order of $\sigma_2\simeq$ 100 fb.
This is a consequence of an intrinsic property of the model. That is, in most 
part of the parameter space, the axial spin-one $Z_2$-boson is more strongly
coupled to fermions than the vector spin-one $Z_1$-boson. This 
peculiarity is shown in Fig.~\ref{fig:a1_a2} for two sample values
of the free $z$-parameter: $z$=0.6, 0.8.
This characteristic will have direct influence on the resonant peaking 
structure of the differential cross section, as we will see in the next 
section.

\begin{figure}[!htbp]
\begin{center}
\unitlength1.0cm
\begin{picture}(7,6)
\put(-4.1,1.8){\epsfig{file=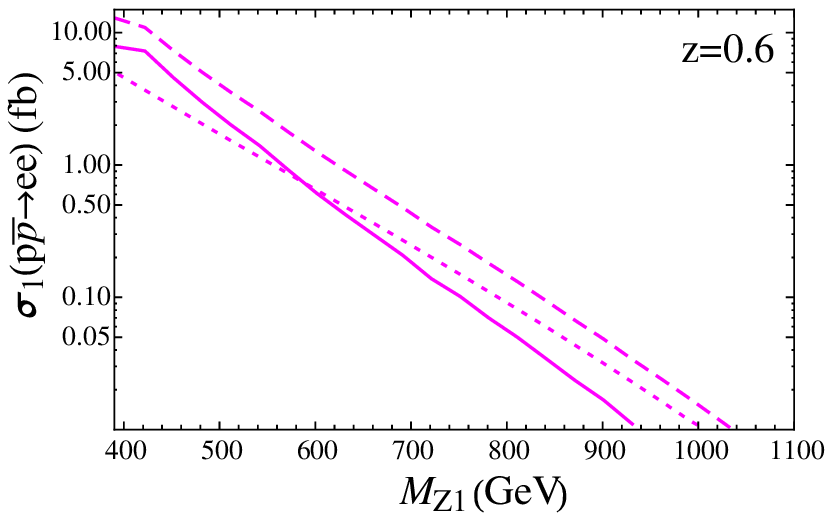,width=7.5cm}}
\put(3.5,1.8){\epsfig{file=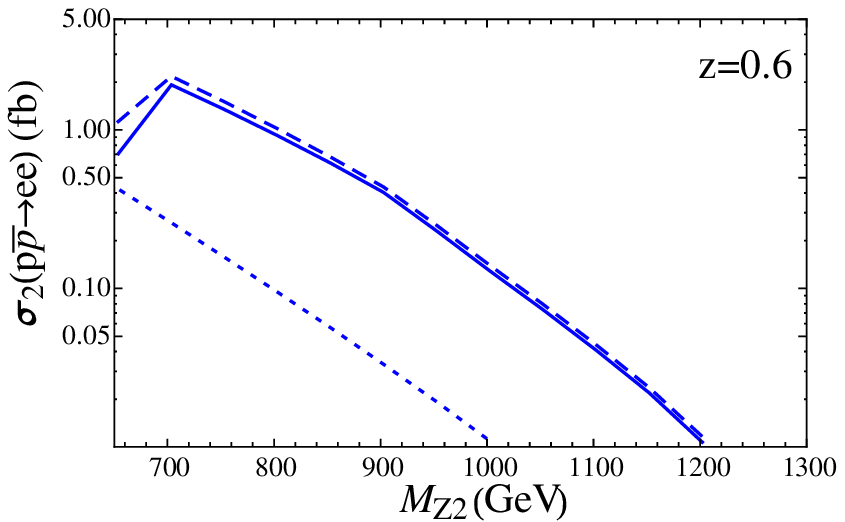,width=7.5cm}}
\put(-4.1,-3.6){\epsfig{file=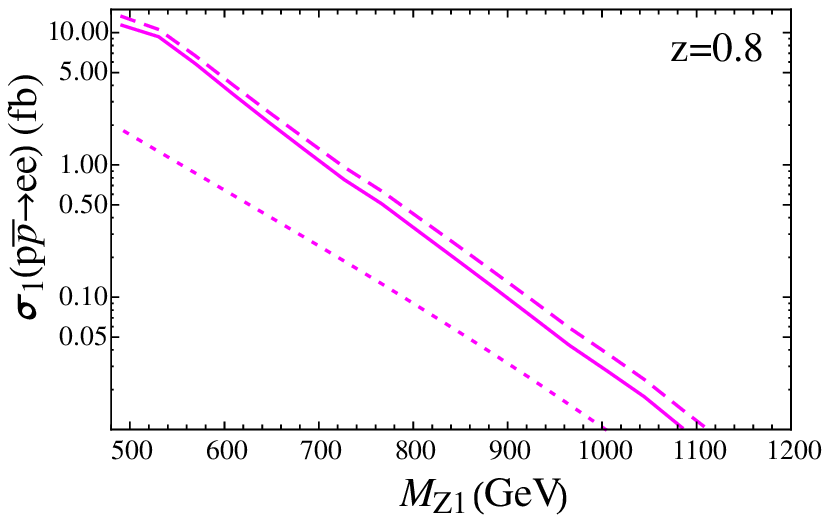,width=7.5cm}}
\put(3.5,-3.6){\epsfig{file=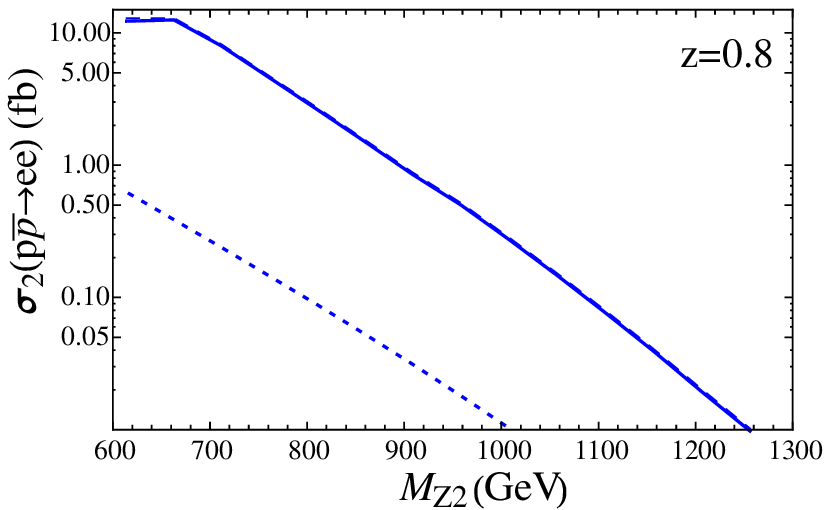,width=7.5cm}}
\put(-4.1,-8.6){\epsfig{file=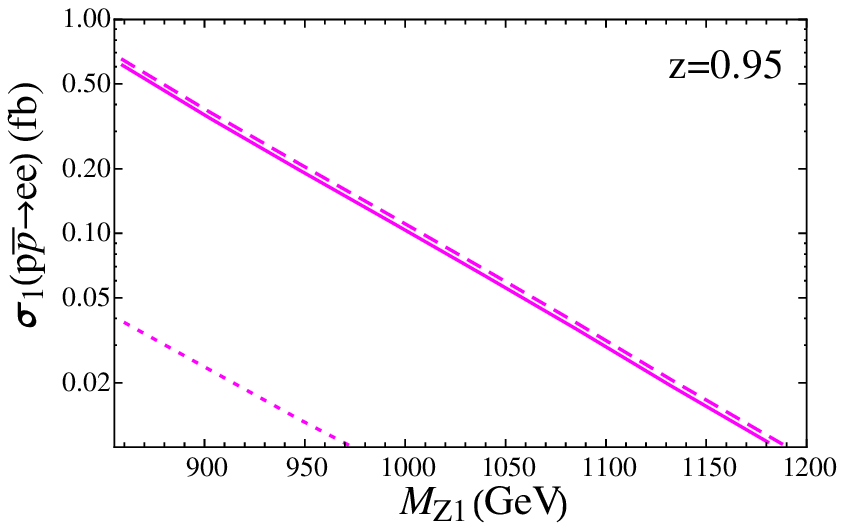,width=7.5cm}}
\put(3.5,-8.7){\epsfig{file=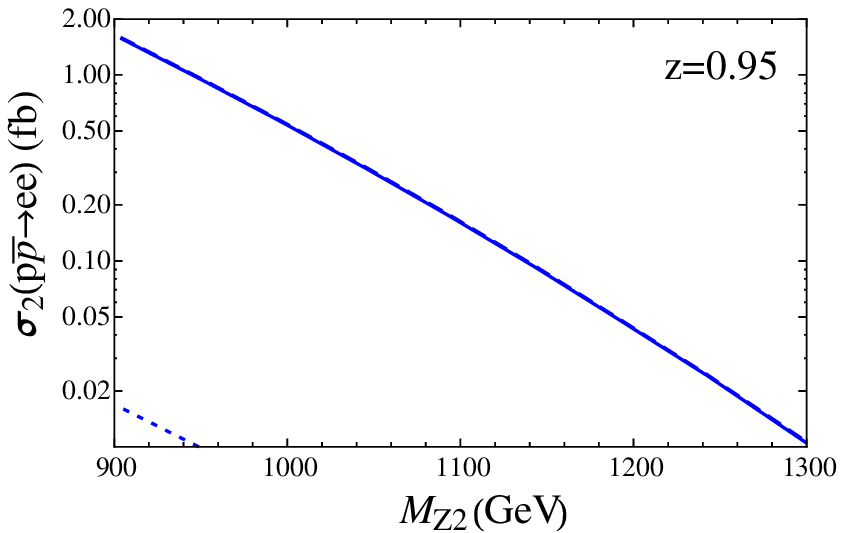,width=7.4cm}}
\end{picture}
\end{center}
\vskip 8.cm
\caption{Left: $Z_1$-boson cross-sections, for the maximal $\hat a^\Pe_{2L}$ 
allowed by EWPT, in the mass window given in Eq.(\ref{window}), as a function 
of $M_{Z_1}$ at the Tevatron. The dashed line represents the total 
cross-section (T), including the interference 
of the $Z_1$-boson signal with both the SM background and the $Z_2$-boson 
signal. The dotted line shows the SM background (B). The solid line 
corresponds to their difference: S=T-B.  
Right: same for the $Z_2$-boson. From top to bottom, the free $z$-parameter 
assumes the three representative values: $z=0.6, 0.8, 0.95$.}
\label{fig:cross-section_Tevatron}
\end{figure}

As comparison, we consider also the $Z_{1,2}$-boson production at the 
Tevatron via the process 
$\Pp\bar\Pp\rightarrow\gamma, Z, Z_{1,2}\rightarrow\Pe^+\Pe^-$. In 
Fig.~\ref{fig:cross-section_Tevatron}, we show the total cross-section under 
the $Z_{1,2}$ resonances as a function of the gauge boson masses, in perfect 
analogy with the LHC plots presented before. At the Tevatron, the expected 
cross section is always below 15 fb.

%\subsubsection{$Z_1$ and $Z_2$ production at the LHC and Tevatron}
%\label{se:neutral_production}
In order to estimate the LHC reach, we now consider in Table 
\ref{tab:scenarios} different choices of mass spectrum.
\begin{table}[h!]
%$\hspace{-0.5cm}$z$=0.6\hspace{4.5cm} $z$=0.8\hspace{4.5cm}$z$=0.95$
\begin{center}
\begin{tabular}{|c|c|c|c|c|}
%\hline $z=0.6$\\
\hline 
$z=0.6$ & $M_{Z_1,Z_2} (\GeV)$ & $\Gamma_{Z_1,Z_2} (\GeV)$ & ${\hat a}^\Pe_{2L}$& ${\hat a}^\Pe_{1L}$\\
\hline \hline
a & 436, 704 & 5.4, 10.8 & 0.33&0.23\\
\hline \hline
b & 552, 903 & 8.8, 23.7 & 0.45& 0.23\\
\hline
\end{tabular}~~~~
\begin{tabular}{|c|c|c|c|c|}
\hline
$z=0.8$ & $M_{Z_1,Z_2} (\GeV)$ & $\Gamma_{Z_1,Z_2} (\GeV)$ & ${\hat a}^\Pe_{2L}$& ${\hat a}^\Pe_{1L}$ \\
\hline \hline
c& 505, 614  & 7.9, 3.3 & 0.29 & 0.33\\
\hline \hline
d& 893, 1107 & 26.6, 23.3 & 0.50& 0.36\\
\hline 
\end{tabular}~~~~\\
\bigskip
\begin{tabular}{|c|c|c|c|c|}
\hline 
$z=0.95$ & $M_{Z_1,Z_2} (\GeV)$ & $\Gamma_{Z_1,Z_2} (\GeV)$ & ${\hat a}^\Pe_{2L}$& ${\hat a}^\Pe_{1L}$\\
\hline \hline
e&889, 968 &16.8,14.2 & 0.60& 0.45\\
\hline \hline
f&1359, 1433 &39.6, 21.6 & 0.64& 0.49\\
\hline
\end{tabular}
\end{center}
\caption{Six representative scenarios for the four-site Higgsless model. In 
the tables, $M_{Z_1,Z_2}$ and $\Gamma_{Z_1,Z_2}$ are physical masses and 
widths of the $Z_{1,2}$-bosons, and ${\hat a}^\Pe_{1L}$, ${\hat a}^\Pe_{2L}$ 
represent the $Z_{1,2}$-boson couplings to the left-handed electron.}
\label{tab:scenarios}
\end{table}
These examples give an idea of the possible scenarios predicted
by the four-site Higgsless model. In the model in fact, as already mentioned, 
the ratio between the gauge boson masses of the first and second triplet, 
$z$, is a free parameter. Hence, the distance between the
two masses is arbitrary as well. We have thus chosen three cases: 
$z=0.6$ representing the flat-metric scenario 
(see \cite{Accomando:2008jh}), giving rise to 
very distant resonances, $z=0.8$ which is a case where the $Z_{1,2}$ bosons 
are closer in mass but still separately measurable, and finally $z=0.95$ 
corresponding to a spectrum which tends to degeneracy 
by approaching the limit $z\rightarrow 1$.

For each $z$-value, we then consider two sets of $M_{Z_1,Z_2}$ masses. The 
first one corresponds to the point in the parameter space, allowed by EWPT 
and by direct Tevatron limits (see next section), which ensures the maximal 
cross-section. And, typically, it coincides with the minimum allowed mass.
The latter set refers instead to the case in which one reaches the 5-event 
threshold in order to claim a discovery at the LHC with $\sqrt{s}=7$ TeV and 
L$=1$ fb$^{-1}$.
We always assume to work with the maximum value of the $Z_2$-boson coupling
 to SM fermions for any given mass and $z$-value. This means that we move 
along the upper part of the contour plots which delimit the area allowed by 
EWPT in the plane ($M_{Z_2}$, ${\hat a}^\Pe_{2L}$), shown in 
Fig. \ref{fig:m2a2l}.

\begin{figure}[!htbp]
\begin{center}
\unitlength1.0cm
\begin{picture}(8,7)
%  \xText{-2.5}{5.3}{\small{(a)~~$M_{1,2}=(500, 1250)\GeV$}}
%  \xText{-1.3}{-0.1}{$\mr{M}_{\inv}(l^+l^-) \normalsize [\GeV]$}
%  \xText{-4.}{3.}{$N_{evt}$}
%  \xText{5.2}{5.3}{\small{(b)~~$M_{1,2}=(1732, 3000)\GeV$}}
%  \xText{6.7}{-0.1}{$\mr{M}_{\inv}(l^+l^-) \normalsize [\GeV]$}
%  \xText{3.9}{3.}{$N_{evt}$}
%  \xText{-2.5}{-0.85}{\small{(c)~~$M_{1,2}=(1000, 1250)\GeV$}}
%  \xText{-1.3}{-6.3}{$\mr{M}_{\inv}(l^+l^-) \normalsize [\GeV]$}
%  \xText{-4.}{-3.3}{$N_{evt}$}
%  \xText{5.2}{-0.85}{\small{(d)~~$M_{1,2}=(1000, 1250)\GeV$}}
%  \xText{6.7}{-6.3}{$\mr{M}_{\inv}(l^+l^-) \normalsize [\GeV]$}
%  \xText{3.9}{-3.3}{$N_{evt}$}
\put(-4.2,0.7){\epsfig{file=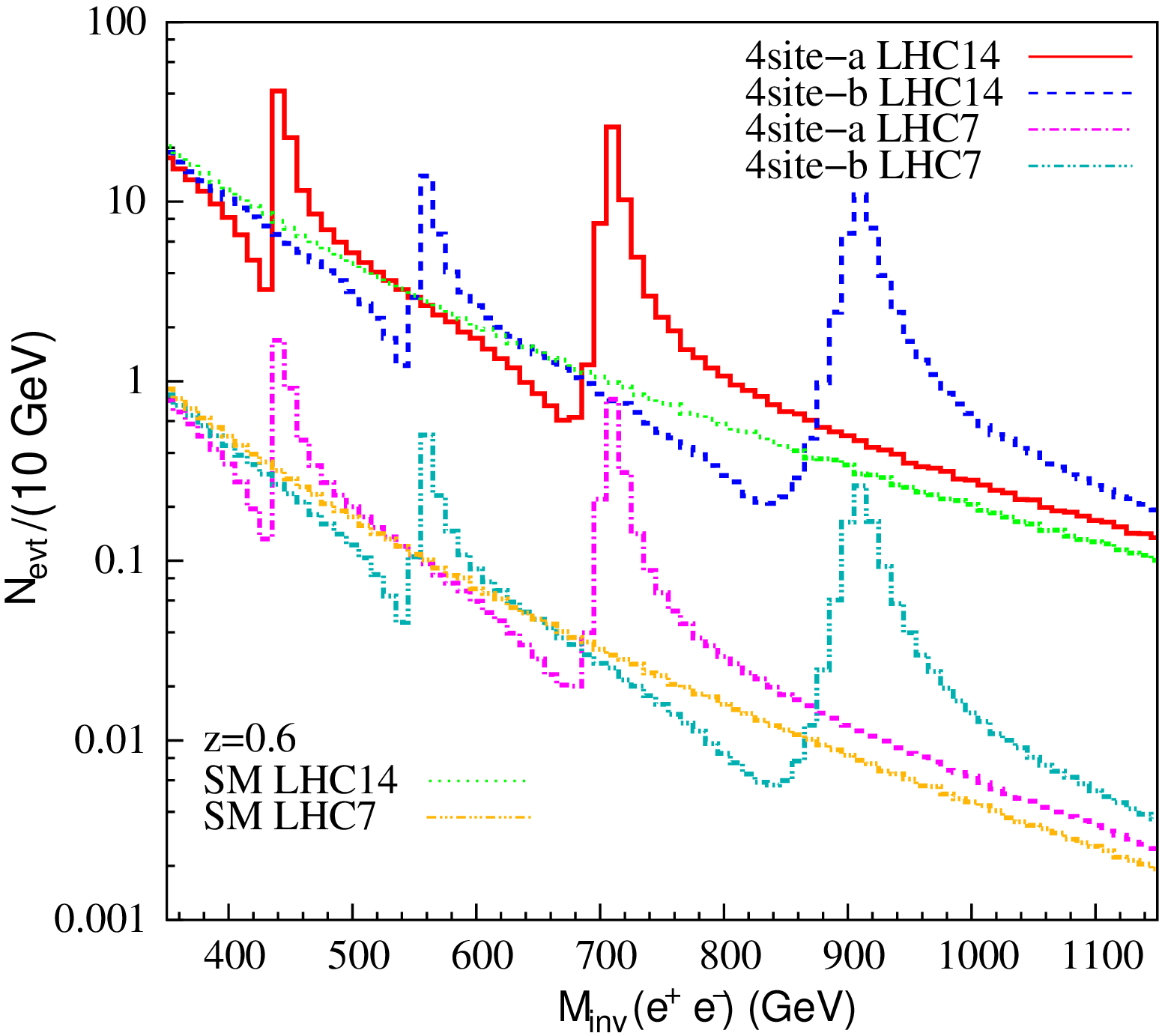,width=7.5cm}}
\put(-4.2,-6.8){\epsfig{file=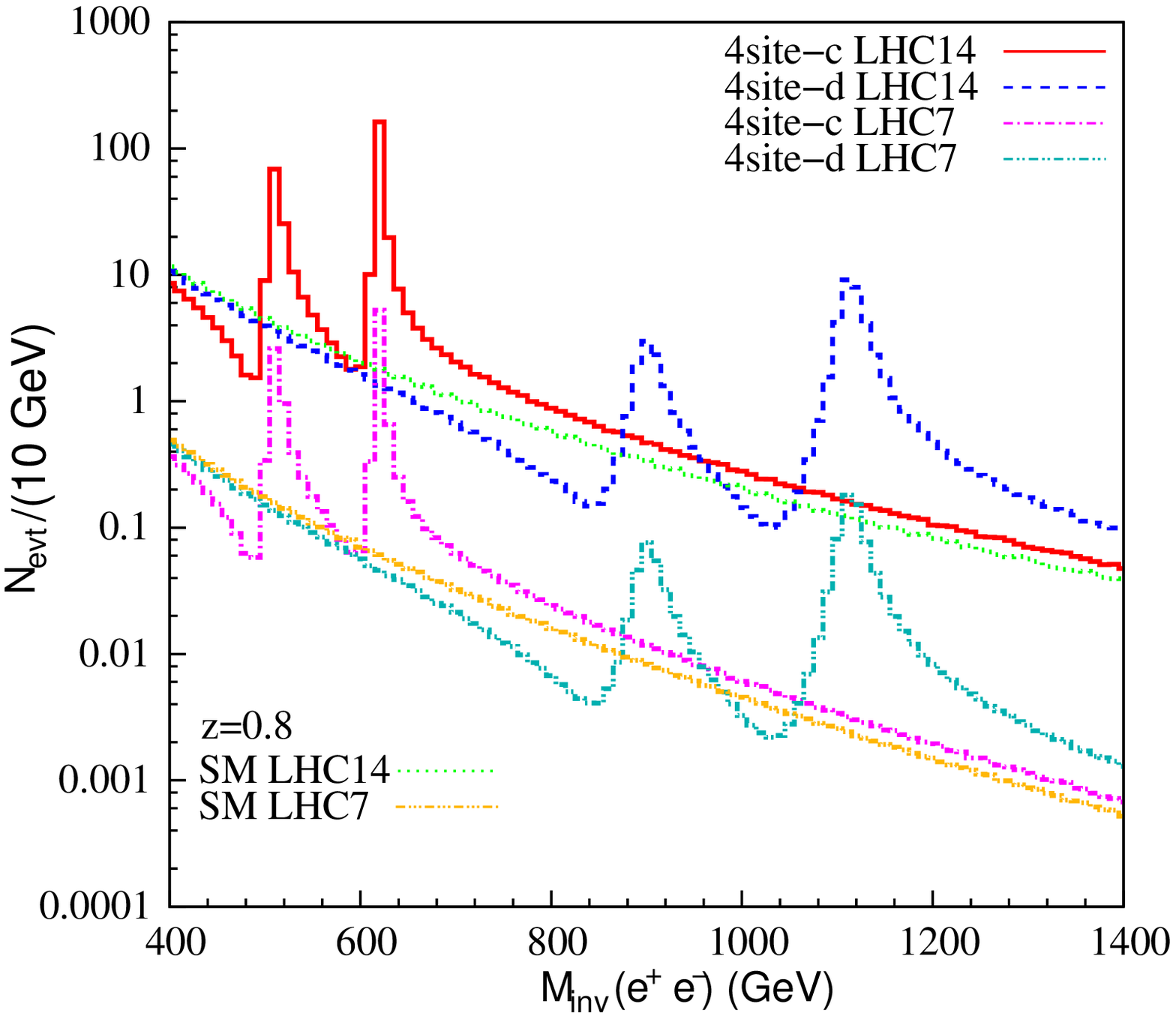,width=8.0cm}}
\put(-4.2,-14.3){\epsfig{file=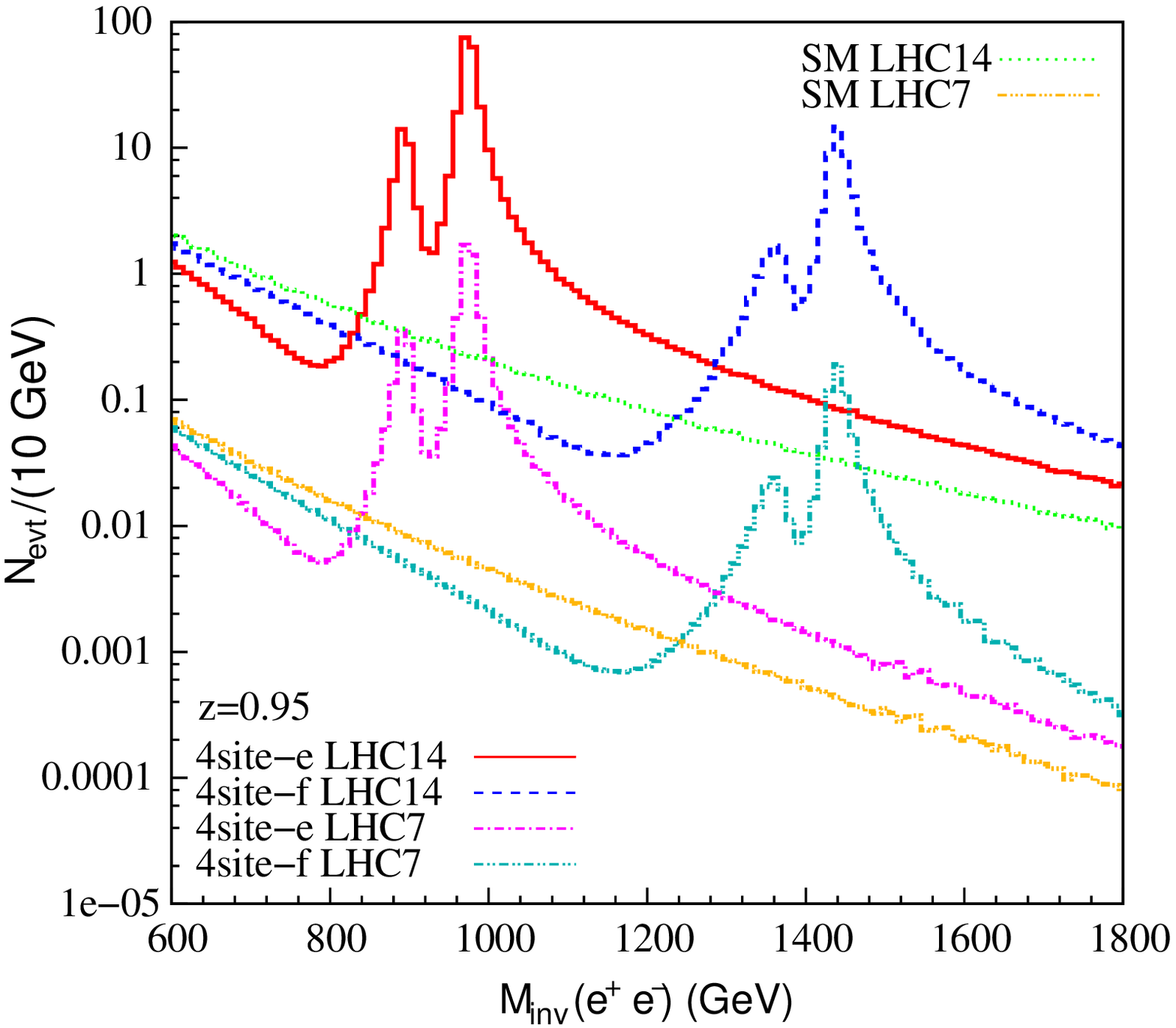,width=8.0cm}}
\put(3.9,0.7){\epsfig{file=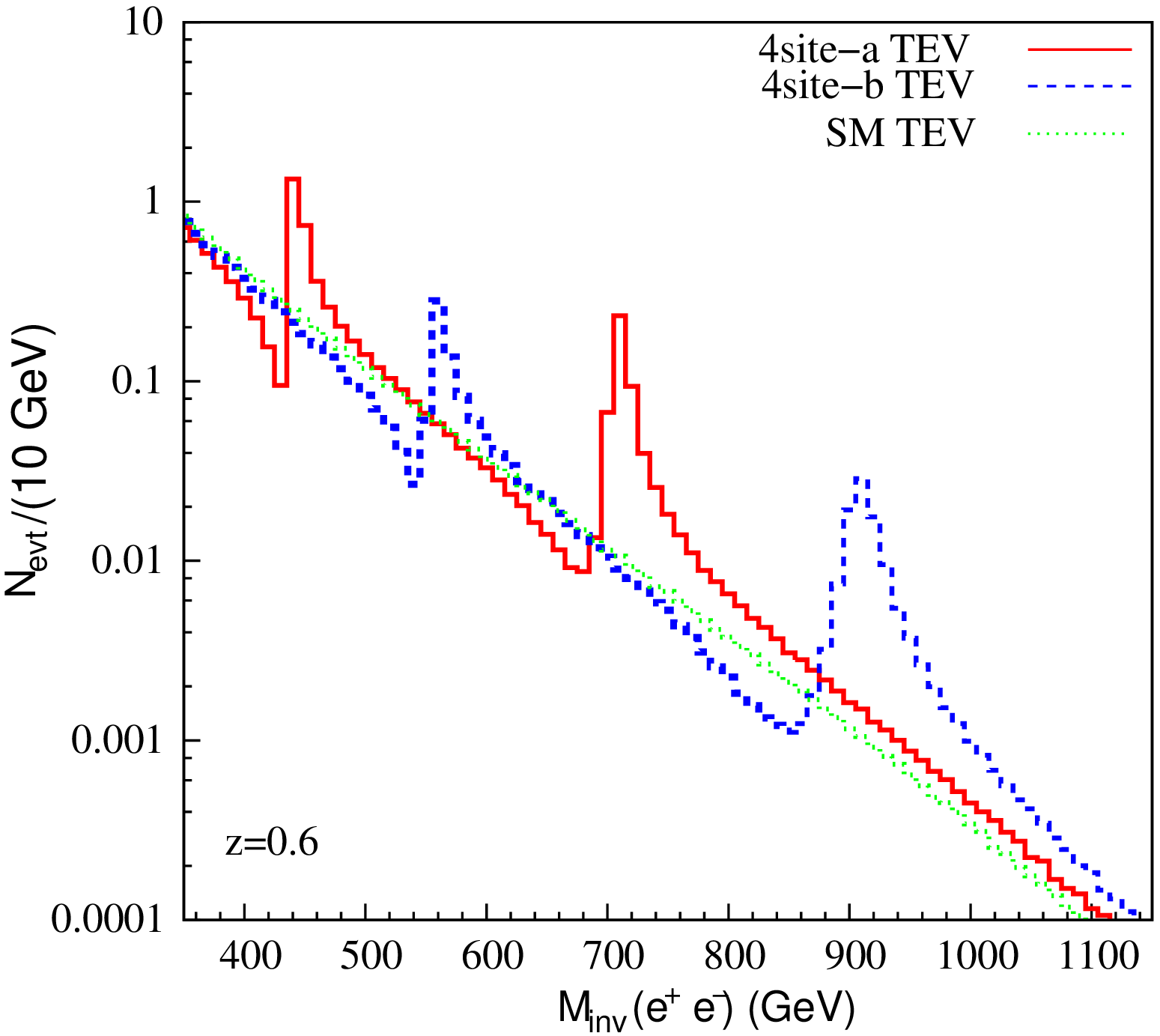,width=7.6cm}}
\put(3.9,-6.85){\epsfig{file=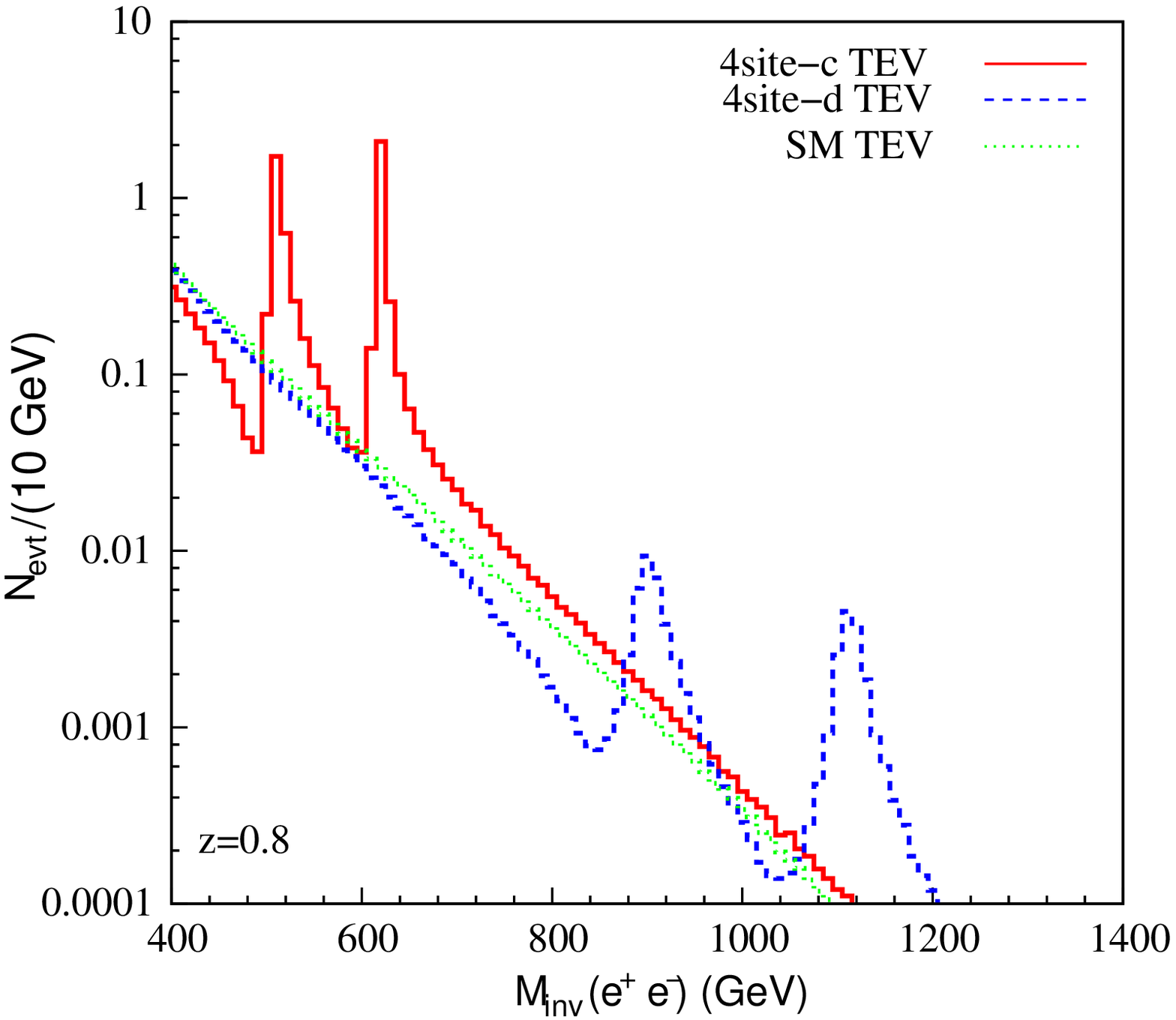,width=8.1cm}}
\put(3.9,-14.3){\epsfig{file=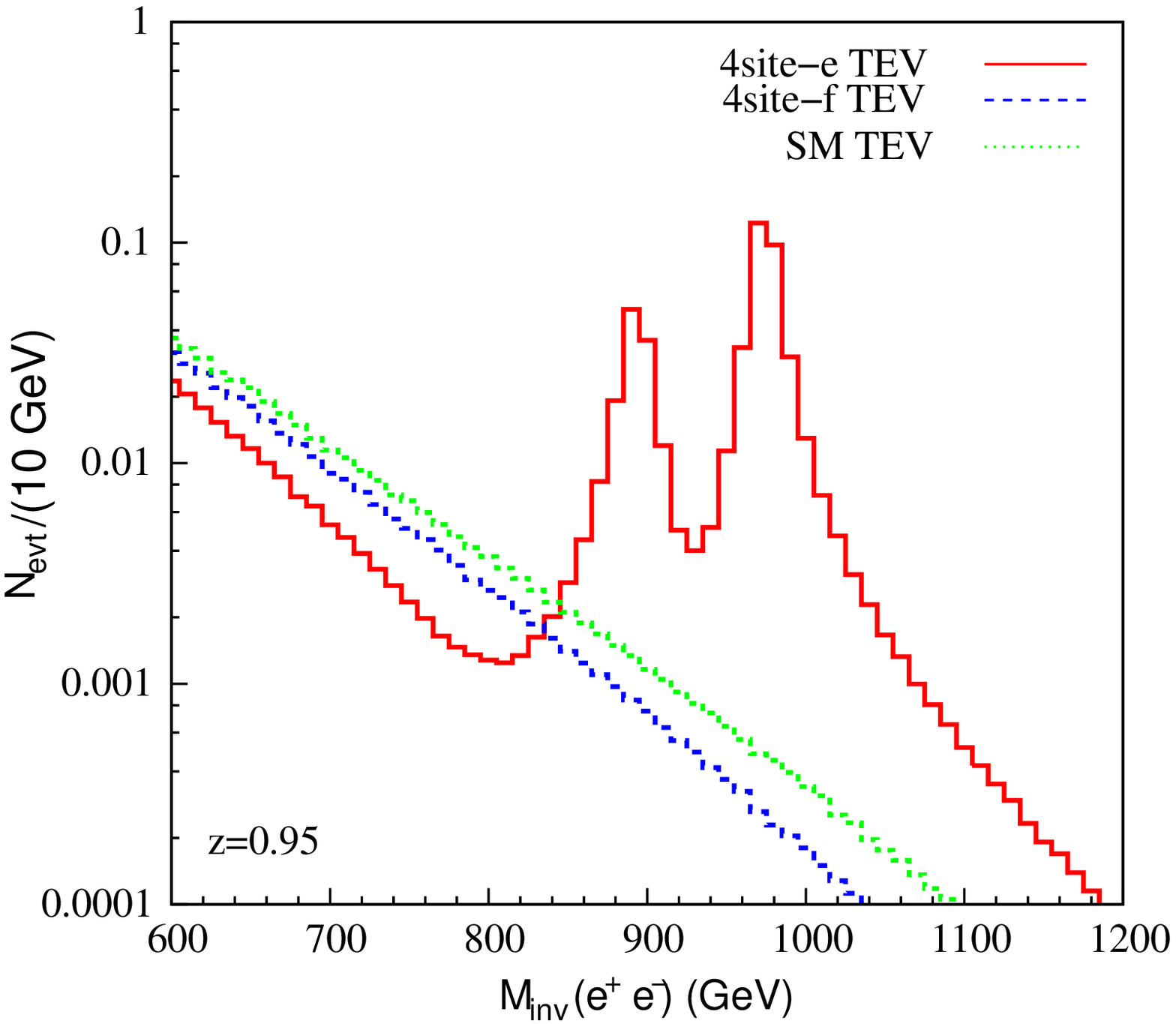,width=7.93cm}}
\end{picture}
\end{center}
\vskip 13.6cm 
\caption{Left: Total number of events over a 10$\GeV$-bin versus
the dilepton invariant mass, $M_{\inv}(\Pe^+\Pe^-)$, for the process
$\Pp\Pp\rightarrow \Pe^+\Pe^-$ at the 7 TeV LHC with L=1 fb$^{-1}$ (lower 
curves) and at the 14 TeV LHC with L=10 fb$^{-1}$ (upper curves), for the  
scenarios given in Table \ref{tab:scenarios}. From top to bottom: $z$=0.6, 0.8, 
and 0.95. Standard cuts and a 90$\%$ detection efficiency for the electron is 
included. Right: same as in the left panel for the process 
$\Pp\bar\Pp\rightarrow \Pe^+\Pe^-$ at the 1.96 TeV Tevatron with 
L=10 fb$^{-1}$. A global 20$\%$ signal acceptance is included.}
\label{fig:minv}
\end{figure}

We are now ready to present some differential cross sections 
for the leptonic process $\Pp\Pp(\Pp\bar\Pp)\to \Pe^+\Pe^-$. 
As an illustration of the behaviour and the impact of the new
predicted particles at the LHC, in the following we analyze the
distribution in the invariant mass of the reconstructed
$Z_{1,2}$-boson, that is $M_{\inv}(\Pe^+\Pe^-)$.
In the previous section, we have seen that there is a lower bound on the 
$Z_{1,2}$-boson masses: $M_{Z_1}>350\GeV$ and $M_{Z_2}>500\GeV$.
We thus select this energy scale by imposing an additional
cut on the invariant mass of the lepton pair, i.e.
$M_{\inv}(\Pe^+\Pe^-)\ge 200\GeV$.
 In Fig.~\ref{fig:minv}, we
plot the total number of events divided by a 10$\GeV$ bin as a function of
the dilepton invariant mass for the six aforementioned scenarios. We have 
checked that all these cases are outside the exclusion limit
from direct searches at the Tevatron with an integrated luminosity
L=3.6$\fba^{-1}$ \cite{D0preliminary} (see next section). 

These examples show four peculiarities of the model. First of all,
the masses of the two neutral resonances $Z_{1,2}$ are not equally
spaced. They can be either very distant for low $z$-values (see
top-left panel in Fig.~\ref{fig:minv}), or they can tend to be
almost degenerate for high $z$-values (see bottom-left panel in 
Fig.~\ref{fig:minv}). For some values of the parameter space (see for example 
the $b$, $d$ and $f$ scenarios), one of the two resonances could also 
disappear leaving a single-resonant spectrum. In this case, the distinctive 
multi-resonant peaking structure of the four-site model would be hidden,  
making the Higgsless model to become highly degenerate with common extra 
U(1) theories predicting just one additional $Z^\prime$-boson.
The same situation would happen for $z=0.95$ (see bottom-left panel in
Fig.~\ref{fig:minv}), if the mass resolution were not small enough to allow a 
separate measurement of the two resonances.

A second feature is related to the interference between signal and
SM background. All  plots exhibit indeed a sizeable depletion of
the total number of events, compared to the SM prediction, in the
off-peak region. A further
distinctive behaviour is represented by the width magnitude. It
indeed increases with the third power of the extra gauge boson mass.
One can thus pass from configurations with very narrow resonances to 
scenarios characterized by broad peaks (or even shoulders). The last feature
concerns the relative size of the $Z_{1,2}$ resonances. In 
Fig.~\ref{fig:a1_a2}, we have indeed shown that in most part of the
parameter space the $Z_1$-fermion couplings are smaller than the
$Z_2$-fermion ones. As a consequence, the height of the
$Z_1$-resonance is less pronounced than the $Z_2$-peak. This feature
can be washed out by the PDF effect, but it is clearly visible in
Fig.~\ref{fig:minv}.

In order to compare the LHC reach at $L=1\fba^{-1}$ with the discovery 
potential of the Tevatron projected at $L=10\fba^{-1}$, in Fig.~\ref{fig:minv} 
(right panel), we plot the $Z_{1,2}$ spectrum for the same setups (only the first setup  of 
the $z$=0.95 scenario is shown in the bottom-right panel). We display  the total number of events over a 
10$\GeV$ bin expected for the two colliders.

To have an idea of the detection rate expected at the LHC with $\sqrt{s}=7$ TeV
 and L=1 fb$^{-1}$ for the Drell-Yan production of the extra $Z_{1,2}$ gauge 
bosons, in Tab.~\ref{tab2} we have listed signal and background event number 
in the two distinct on-peak regions
$|M_{\inv}(\Pe^+\Pe^-)-M_{Z_1,Z_2}|<max(\Gamma_{Z_1,Z_2}/2, R_{LHC})$ for the 
six considered scenarios. The signal event number is calculated as the 
difference between the total number of events and the background.
The Tevatron expected rates for the same six setups, calculated for 
10 fb$^{-1}$ with the same procedure, turn out to be much smaller and only 
the case $c$ gives sizeable statistical significance ($\sigma(Z_1)=$11.7 and
$\sigma(Z_2)=26.0$).

\begin{table}[!htb]
\begin{center}
\begin{tabular}{|c|c|c|c|c|c|c|c|c|}
\hline &$M_{Z_1,Z_2} (\GeV)$ & $\Gamma_{1,2}(\GeV)$ & $N_{\evt}^{\sig}(Z_1)$ &
$N_{\evt}^{\backg}(Z_1)$ & $\sigma (Z_1)$ & $N_{\evt}^{\sig}(Z_2)$ &
$N_{\evt}^{\backg}(Z_2)$
& $\sigma (Z_2)$ \\
\hline \hline
a&436,704 &5.4,10.8& 17 & 5 & 7.6 & 12 & 1 & 12.0 \\
\hline
b&552,903 &8.8,23.7& 5 & 2 & 3.5 & 6 & 1 & 6.0 \\
\hline
c&505,614 &7.9,3.3& 31 & 3 & 17.9 & 59 & 1 & 59.0 \\
\hline
d&893,1107 &26.6,23.3& 1 & 1 & 1.0 & 5 & 1 & 5.0 \\
\hline
e&889,968 &16.8,14.2 & 8 & 1 &8.0 & 39 & 1 & 39.0\\
\hline
f&1359,1433 &39.6,21.6& 1 & 1 & 1.0 & 5 & 1 & 5.0\\
\hline
\end{tabular}
\end{center}
\caption{The first three columns represent the scenario. The next
three columns give signal (including the interference with the SM background)
and SM background event number for the $Z_1$ production, and the statistical 
significance $\sigma =N_{\evt}^{\sig}/\sqrt{N_{\evt}^{\backg}}$ at the LHC 
with $\sqrt{s}=7$ TeV and L=1 fb$^{-1}$. The last three  columns show the same
results for the $Z_2$ production.} \label{tab2}
\end{table}

\section{$Z_{1,2}$ exclusion and discovery reach at the Tevatron and the LHC}
\label{disco}

In this section, we discuss the prospects of discovering the two neutral
spin-1 bosons predicted by the four-site Higgsless model at the Tevatron and 
LHC.

Let us start by deriving the present exclusion limits on $Z_{1,2}$-bosons
from the Tevatron experiment. We take as a reference the recent preliminary 
analysis of the neutral DY-channel into electron-pairs performed by the D0 
collaboration at the collected luminosity L=3.6 fb$^{-1}$ 
\cite{D0preliminary}. There, observed and expected 95$\%$ C.L. upper limits 
on $\sigma(\Pp\bar \Pp\to Z')\times Br(Z'\to \Pe\Pe)$ are derived as a 
function of $M_{Z^\prime}$. 
%The expected bound is computed from the Monte Carlo estimated SM background. 
More in detail, the D0 95$\%$ C.L. sensitivity to an hypotetical 
$Z^\prime$-boson has been derived by calculating the DY cross section upper 
bound within an asymmetric mass window around the x-axis M-value, i.e. 
$\M_{\inv}(\Pe^+\Pe^-)\ge\M-3\R_{\TEV}$ where $\R_{\TEV}\simeq 3.4\%\M$ is 
the approximated D0 mass resolution (see Ref. \cite{D0preliminary} and 
references therein). A global signal acceptance of about 20$\%$ is included 
(its precise value depends on the energy scale as described in 
Ref. \cite{D0preliminary}). This allows one to extract direct limits
on the $Z^\prime$-mass from the purely deconvoluted, uncut, theoretical DY
cross section within any given extra $U(1)$ model.

Following the D0 {\it strategy}, the limit on the four-site model has
been obtained by computing the DY cross section in the interval
$\M_{\inv}(\Pe^+\Pe^-)\ge\M_{\PZ_1}-3\R_{\TEV}$, in order to include the
effect of both $\PZ_{1,2}$ resonances, and assuming for each
$\M_{\PZ_1}$-mass the maximal ${\hat a}^\Pe_{2L}$-coupling allowed by EWPT.
No cuts and no detection efficiency are included. The mass limit on the
lightest $\PZ_1$-boson is extracted from the intersection of the theoretical
cross section and the 95$\%$ C.L. upper bound derived from the observed data.
The four-site model is much less constrained compared to other popular 
$\PZ^\prime$ theories owing to generally smaller couplings between SM fermions 
and extra gauge bosons. Assuming maximal values for the $\PZ_{1,2}$-boson 
couplings to SM fermions, the direct mass limit is in fact given by: 
$\M_{\PZ_1}\ge 520\GeV$ for $z$=0.8. For smaller $z$-values, EWPT give very 
strong bounds on $\hat a^\Pe_{2L}$ in the low mass region so that Tevatron is 
not effective. For higher $z$-values, the mass range allowed by the 
$x$-approximation is out of the present Tevatron reach.

\begin{figure}[!htbp]
\begin{center}
\vspace{-.8cm}
\unitlength1.0cm
\begin{picture}(8,7)
\put(-3.3,0.6){\epsfig{file=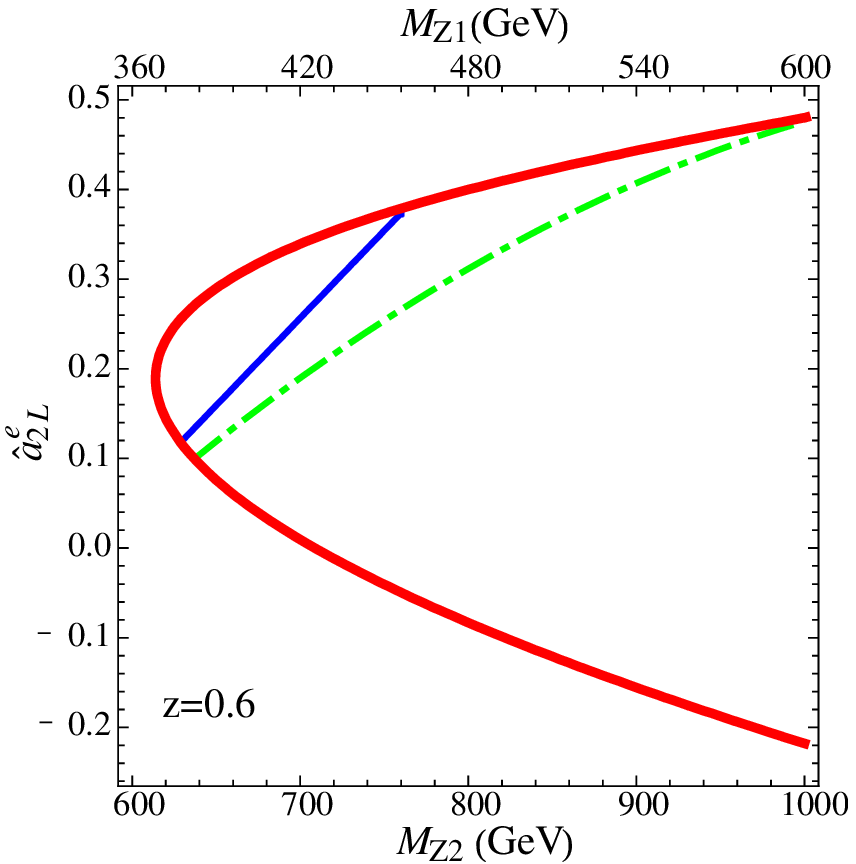,width=6.6cm}}
\put(4.0,0.6){\epsfig{file=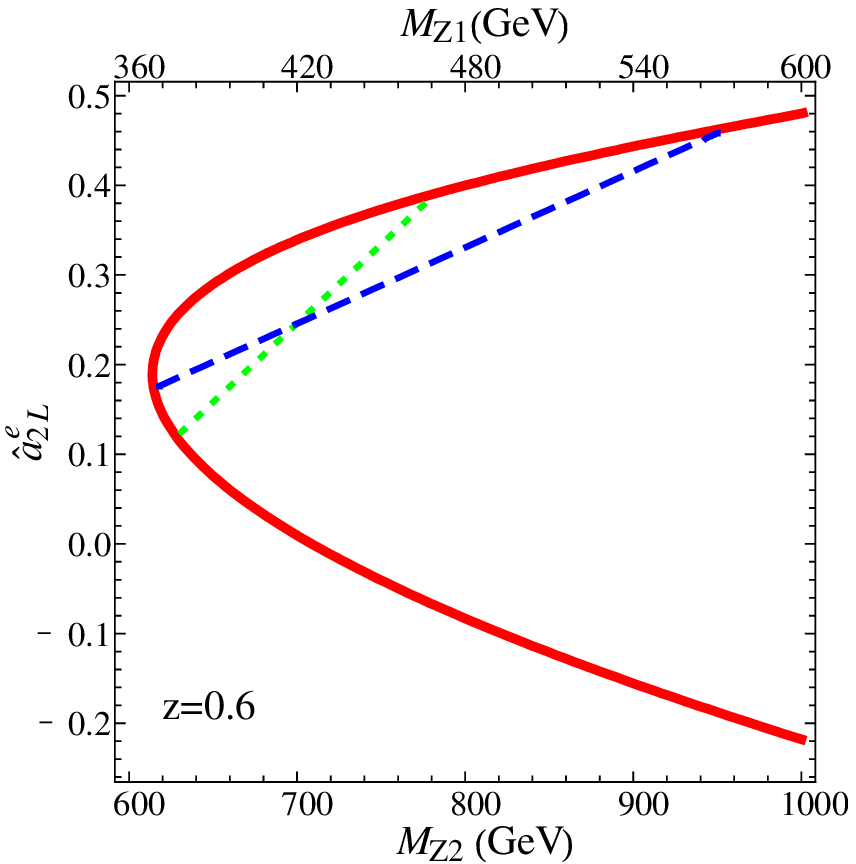,width=6.6cm}}
\put(-3.3,-6.3){\epsfig{file=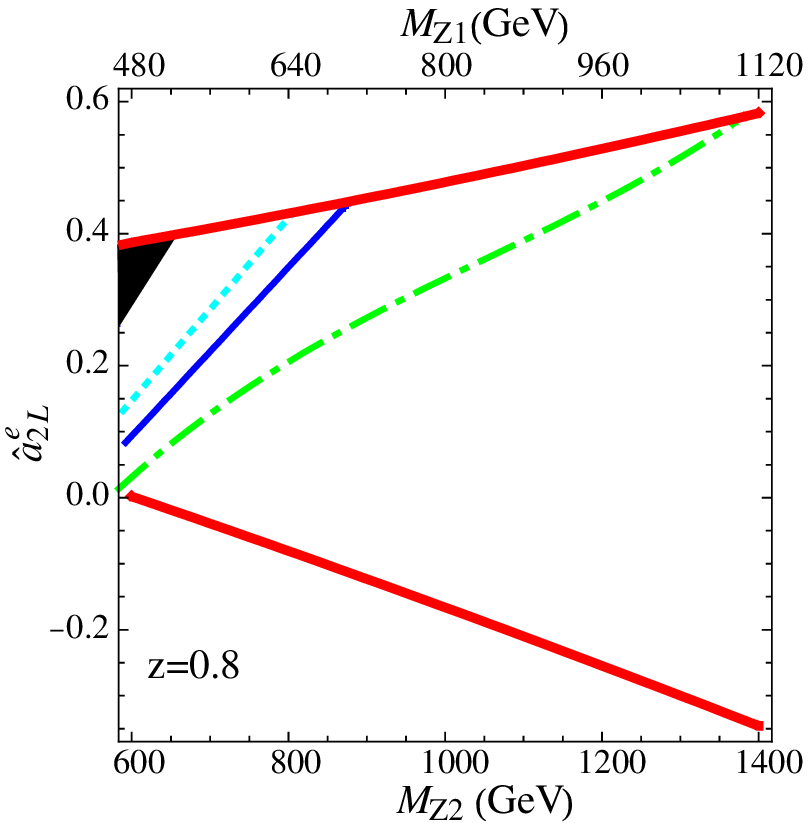,width=6.6cm}}
\put(4.0,-6.3){\epsfig{file=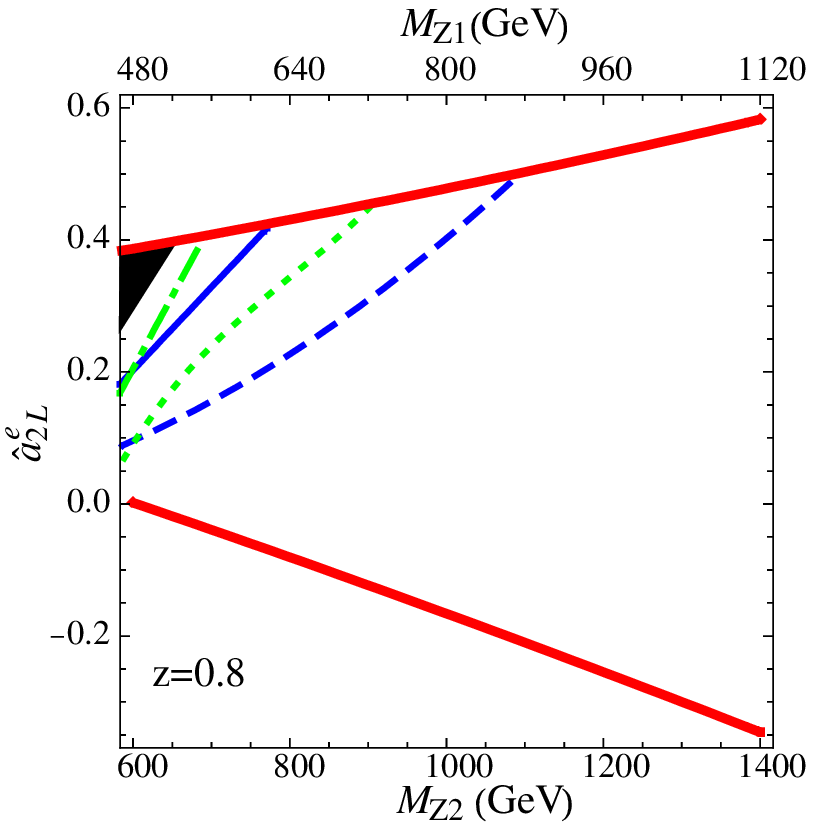,width=6.6cm}}
\put(-3.3,-13.2){\epsfig{file=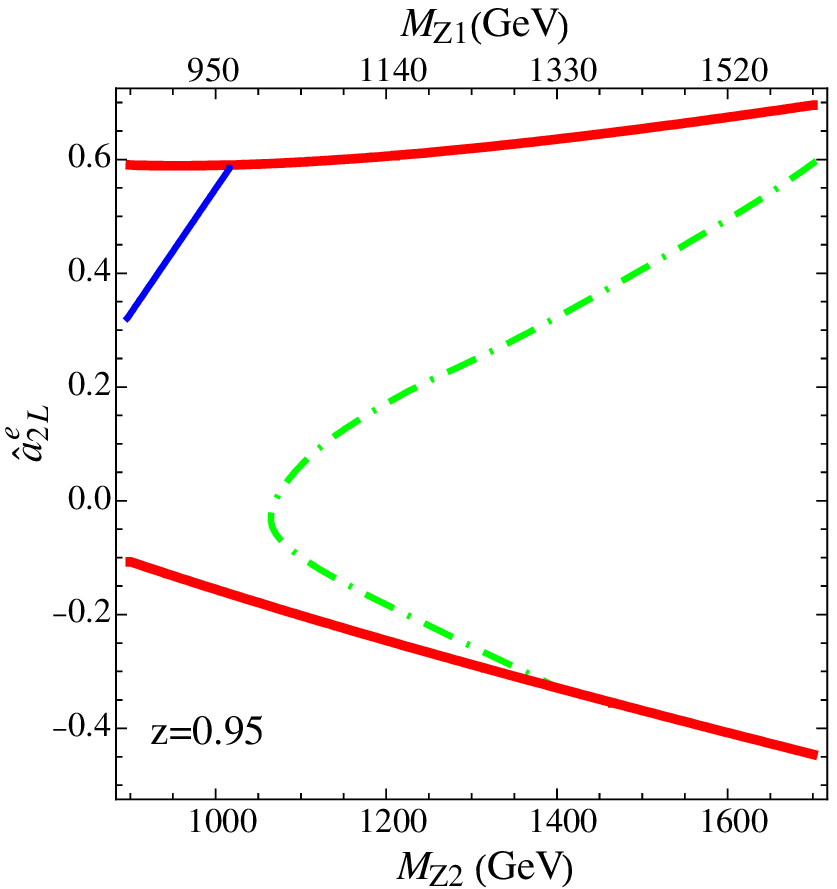,width=6.3cm}}
\put(4.0,-13.2){\epsfig{file=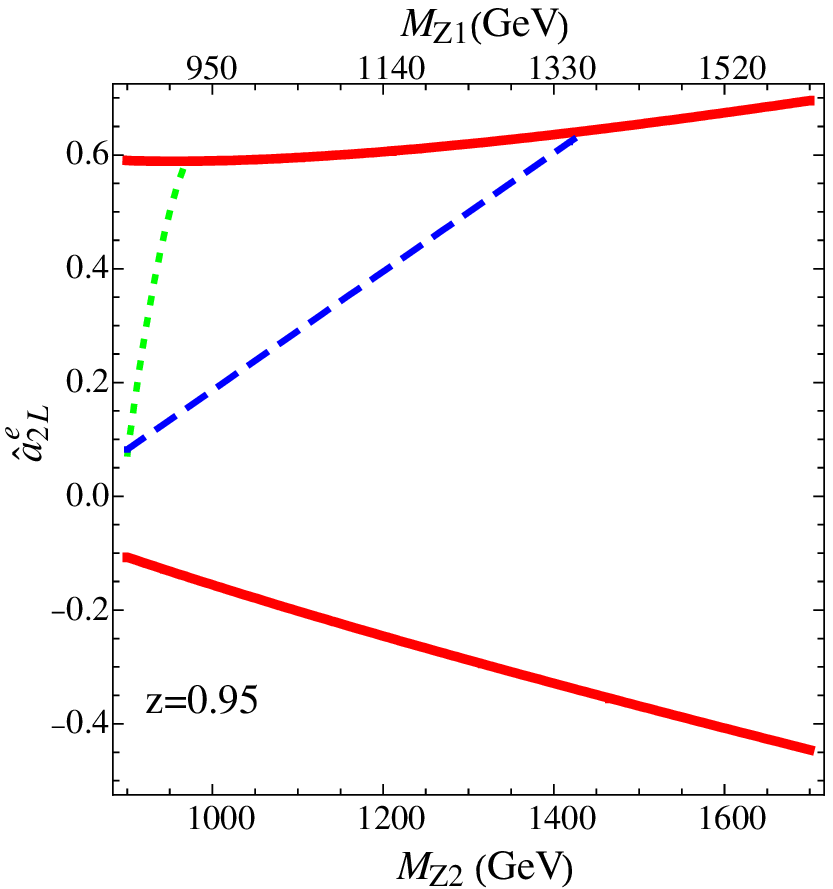,width=6.3cm}}
\end{picture}
\end{center}
\vskip 12.7cm \caption{Left: Exclusion at the Tevatron with L=10 fb$^{-1}$
(solid blue line), and at the 7$\TeV$ LHC with L=1 fb$^{-1}$ (dot-dashed green
line). 
The black triangle represents the exclusion limit from recent preliminary D0 
data taken at L=3.6 fb$^{-1}$. As comparison, the dashed cyan line shows the 
expected exclusion limit from the Tevatron with 3.6 fb$^{-1}$.
Right: $\PZ_1$-boson discovery (dot-dashed green line) and $\PZ_2$-boson
discovery (solid blue line) at the Tevatron with L=10 fb$^{-1}$. Also shown,
the $\PZ_1$-boson discovery (green dotted line) and the $\PZ_2$-boson
discovery (blue dashed line) at the 7$\TeV$ LHC with L=1 fb$^{-1}$. From top to
bottom: $z$=0.6, 0.8, 0.95.  The thick red solid lines give the allowed parameter space from
EWPT and unitarity.}
\label{excldiscov}
\end{figure}

Starting from the above-mentioned limit on the $\PZ_{1,2}$-boson mass
at fixed $z$-value, we have extended the D0 exclusion criteria to the full
parameter space allowed by EWPT for the three reference choices of the $z$ parameter: $z=$0.6, 0.8 and 0.95. The result is
represented by an excluded region in the conventional plane
$( M_{Z_2},\hat{a}_{2L}^\Pe)$. In Fig. \ref{excldiscov}, for $z=0.8$ the 
portion of the parameter space excluded by the present D0 data is depicted as 
a black triangle area (for the other choices of the $z$ parameter the present 
Tevatron data give no restrictions).

As a useful statistical tool, we have moreover verified that the exclusion 
contour in the plane $(M_{Z_2},\hat{a}_{2L}^\Pe)$ at fixed $z$-value that
is obtained by intersecting the theoretical cross section with the 95$\%$ C.L.
upper bound coming from the expected SM background coincides, with good
accuracy, with the exclusion contour derived by using the Poisson statistical
indicator \cite{Bityukov:2000tt}. The agreement is within a few per mill once 
the D0 signal acceptance is included in the evaluation of the Poisson 
distribution.
When discussing future prospects of discovering or excluding the four-site
Higgsless model at Tevatron and LHC, we will thus rely on the Poisson
statistics.

In the following, we compare the $\PZ_{1,2}$-boson exclusion and discovery
reach at the upgraded Tevatron with luminosity L=10 fb$^{-1}$, and at the
7$\TeV$ LHC with L=1 fb$^{-1}$. We assume the signal acceptance setups
described in Sec. \ref{se:setup}.

In the left panel of Fig.~\ref{excldiscov}, we show the exclusion contour
plots in the plane $( M_{Z_2},\hat{a}_{2L}^\Pe)$ for three values of the free
$z$-parameter: $z$=0.6, 0.8, 0.95 from top to bottom. The black triangle
represents the exclusion region based on recent preliminary D0 data, as 
explained above. The solid blue line indicates the 2$\sigma$-exclusion 
potential that
Tevatron should reach in the next two years, i.e. assuming a project
luminosity L=10 fb$^{-1}$. Finally, the dot-dashed green line gives the
2$\sigma$-exclusion reach at the 7$\TeV$ LHC, assuming an integrated
luminosity L=1~fb$^{-1}$. In absence of data, these contour plots have
been computed by making use of the Poisson distribution, as mentioned above.
They have been derived by applying the D0 counting strategy previously
described, i.e. by integrating the cross section in the domain
$\M_{\inv}(\Pe^+\Pe^-)\ge\M_{\PZ_1}-3\R_{\TEV}$, and taking into account the
signal acceptance given in Sec. \ref{se:setup}.
While in the next two years Tevatron could exclude the four-site model up to
energy scales of the order of $M_{\PZ_2}\gtrsim$ 750, 880, 1050 GeV, for 
maximal $\hat{a}_{2L}^\Pe$ couplings allowed by EWPT and 
$z=$0.6, 0.8, 0.95 respectively, the early stage of the
LHC could extend this range to the mass limit 
$M_{\PZ_2}\gtrsim$ 950, 1400 GeV for $z=$0.6, 0.8 respectively and will
 cover the 
whole mass range allowed by the unitarity bound for $z=$0.95.

In the right panel of Fig.~\ref{excldiscov}, we show instead the discovery
contour plot in the plane $( M_{Z_2},\hat{a}_{2L}^\Pe)$ for $z$=0.6, 0.8, 0.95
from top to bottom. As for the left panel, the black triangle represents the
exclusion region based on recent preliminary D0 data.
The procedure we apply in order to derive the discovery reach at the Tevatron
and the LHC is different from the D0 counting strategy adopted for the 
exclusion contour plots. In this case, we aim in fact to distinguish the 
$\PZ_1$ and $\PZ_2$-resonances. Hence, we pursue a shape reconstruction 
analysis. We
thus calculate the cross section below each resonant peak, taking into
account signal, SM background and their mutual interference. Following Ref.
\cite{Basso:2010pe}, we select the mass window
$|\M_{\inv}(\Pe^+\Pe^-)-\M_{\PZ_1,\PZ_2}|\le max (0.5\times
\Gamma_{\PZ_1,\PZ_2}, \R)$ where $\R$ is the mass resolution at Tevatron
($\R_{\TEV}$) or LHC ($\R_{\LHC}$), and we include the signal acceptance
as given in Sec. \ref{se:setup}. By applying the Poisson statistical method,
we obtain the $\PZ_{1,2}$-boson discovery potential shown in the right panel
of Fig.~\ref{excldiscov}.
The gree dot-dashed (blue solid) line indicates the $\PZ_1$ ($\PZ_2$) 
discovery contour
that Tevatron should be able to cover in the next two years, i.e. assuming a
project luminosity $L=$10 fb$^{-1}$. As a comparison, the green dotted 
(blue dashed) line shows the $\PZ_1$ ($\PZ_2$) discovery reach at the 
7 TeV LHC, assuming
an integrated luminosity $L$=1 fb$^{-1}$.

\begin{figure}[!htbp]
\begin{center}
\unitlength1.0cm
\begin{picture}(8,7)
\put(-3.3,0.6){\epsfig{file=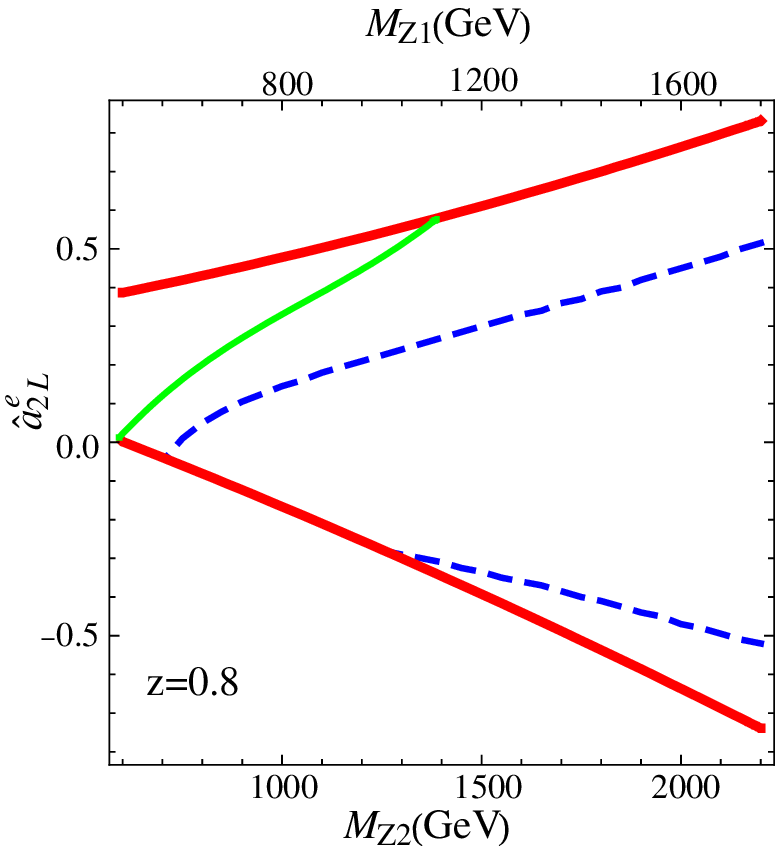,width=6.6cm}}
\put(4.0,0.6){\epsfig{file=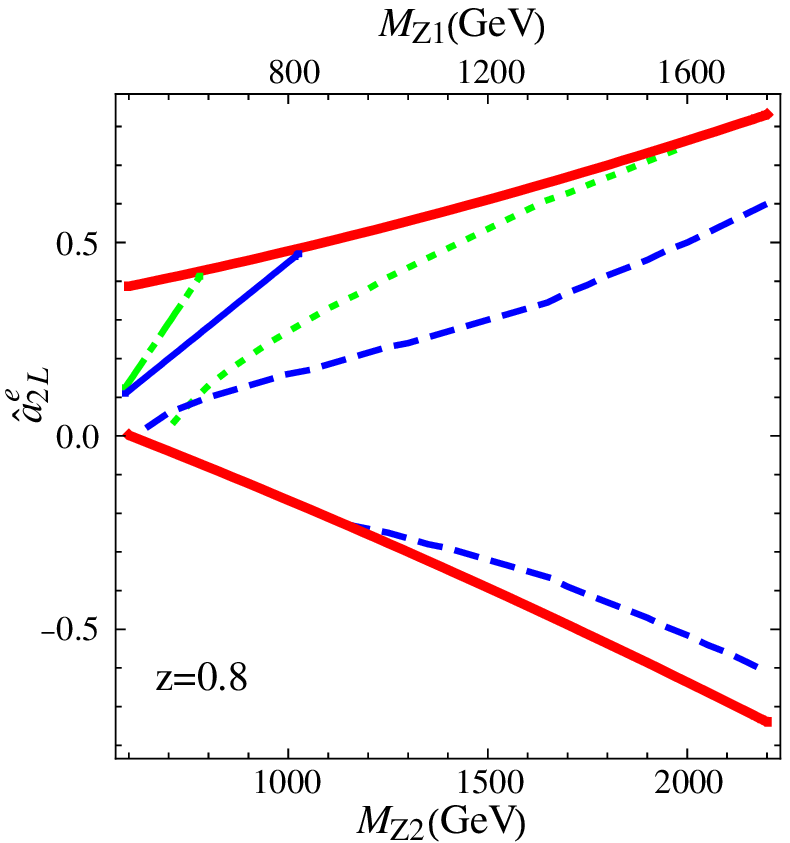,width=6.6cm}}
\end{picture}
\end{center}
\vskip -1.cm \caption{Left: Exclusion at the 7$\TeV$ LHC with L=1 fb$^{-1}$ 
(solid green line), and at the 14$\TeV$ LHC with L=10 fb$^{-1}$ (dashed blue 
line). Right: $\PZ_1$-boson discovery (dot-dashed green line) and $\PZ_2$-boson
discovery (solid blue line) at the 7$\TeV$ LHC with L=1 fb$^{-1}$. Also shown,
as comparison, the $\PZ_1$-boson discovery (green dotted line) and the 
$\PZ_2$-boson discovery (blue dashed line) at the 14$\TeV$ LHC with 
L=10 fb$^{-1}$. All curves have been derived for $z$=0.8. The thick red solid lines 
give the allowed parameter space from EWPT.}
\label{excldiscov_14tev}
\end{figure}

As one can see, independently on the collider, the $\PZ_2$-boson discovery
potential is enhanced compared to the $\PZ_1$-boson. This is due to the
nature of the two spin-1 particles in the four-site model, which has peculiar
consequences on the structure of the $\PZ_{1,2}$-boson couplings to the
ordinary matter (see Fig.~\ref{fig:a1_a2}).
Still, there are regions in the parameter space where the two resonances
could be simultaneously observed. For example, for $z=$0.8 up to energy scales of the order of
$M_{\PZ_2}\simeq$700 (900) GeV at the Tevatron (LHC), the four-site Higgsless
model could manifest its distinctive multi-resonance signature. Beyond those
scales, only the $\PZ_2$-boson could be visible giving rise to the well known
degeneracy problem which affects all extra U(1) theories.
Leaving apart these problematics, which are beyond the scope of this paper,
the global discovery reach during the early run of the LHC is substantial.
The $\PZ_2$-boson could be detected up to $M_{\PZ_2}\lesssim$ 950, 1100, 1400 GeV at the
7$\TeV$ LHC with $L=$1 fb$^{-1}$ for $z=$0.6, 0.8, 0.95 respectively. Owing to the poor signal acceptance, the
Tevatron with $L=$10 fb$^{-1}$ is not expected to go beyond
$M_{\PZ_2}\le$ 770 GeV for $z=$0.8 and it is not effective for $z=$0.6, 0.95.
Looking at the future, in Fig.~\ref{excldiscov_14tev} we show exclusion and
discovery contours at the 14 TeV LHC with L=10 fb$^{-1}$. We consider the
same signal acceptance as in Sec. \ref{se:setup}. We take as representative
example the case $z$=0.8 and we plot also a corresponding contours for the 7 TeV LHC for comparison. We see 
that the second stage of LHC could show the two resonance signature in the electron channel of the Drell-Yan process 
up to $M_{Z_2}=2$ TeV (for $z=0.8$).  Notice that in our previous analysis \cite{Accomando:2008jh} we did not include
any reconstruction efficiency and we summed over electrons and muons.  
In this work we focused only on the electron channel,
 following the D0 data analysis on high mass neutral resonance search in Drell-Yan process. Similar limits
are derived by the CDF collaboration using the muon channel. The two channels 
could be combined, in order to improve the statistics, taking into account 
the different efficiencies \cite{Basso:2010pe}.

\begin{figure}[!htbp]
\begin{center}
\unitlength1.0cm
\begin{picture}(8,7)
\put(-4.3,0.6){\epsfig{file=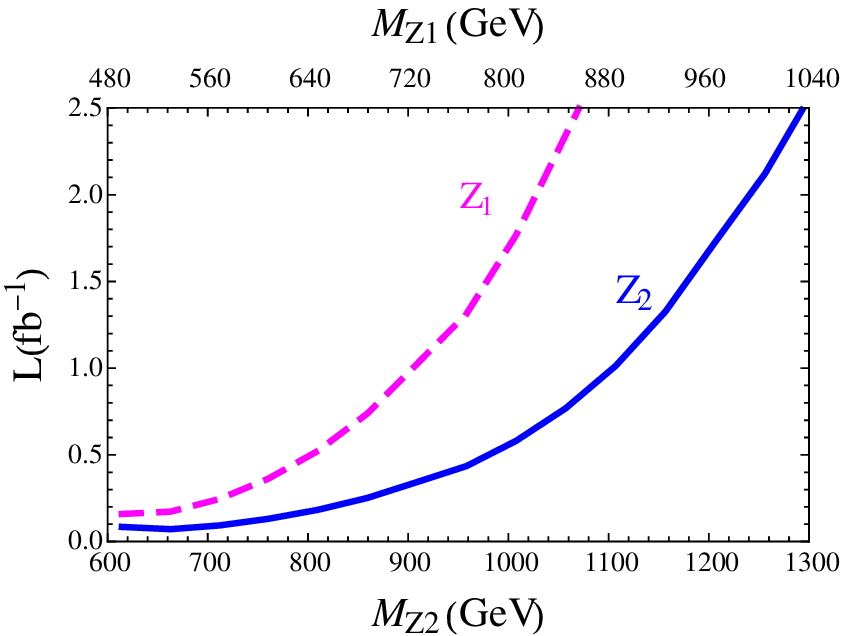,width=8cm}}
\put(4.0,0.6){\epsfig{file=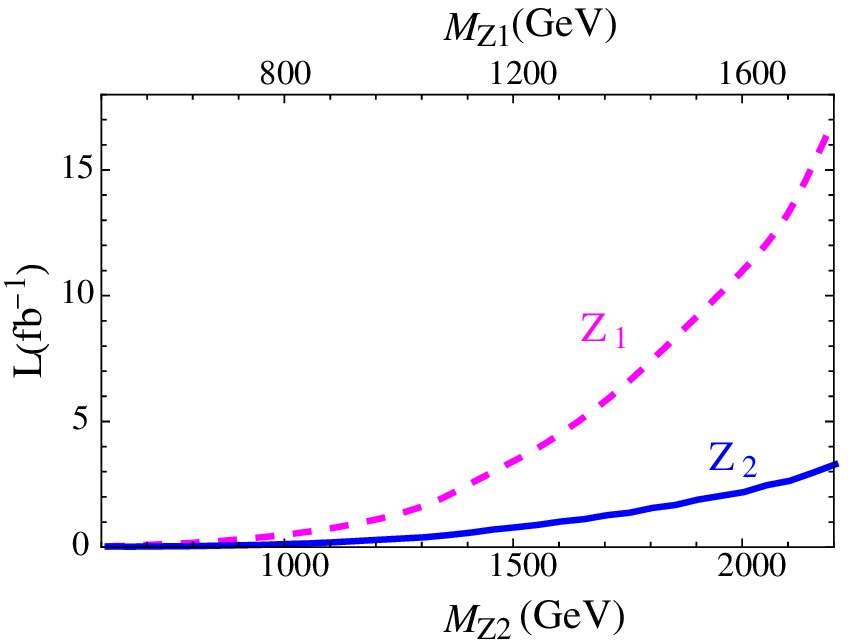,width=8cm}}
\end{picture}
\end{center}
\vskip -1.cm \caption{Left: Minimum luminosity needed for a
5$\sigma$-discovery of $\PZ_1$ (dashed magenta line) and $\PZ_2$ (solid blue
line) at the 7 TeV LHC, assuming maximal $\PZ_2$-boson couplings to SM
fermions and $z$=0.8. Right: same curves at the upgraded 14 TeV LHC.}
\label{disc_lumi}
\end{figure}

To conclude, for the same case $z$=0.8, we show in Fig.~\ref{disc_lumi} the
minimum luminosity needed to claim a $\PZ_{1,2}$-boson discovery during the
early stage (left panel) and project stage (right panel) of the LHC.
While the 7 TeV LHC would need very high luminosity (L$\simeq \mbox{200 fb}^{-1}$) to 
cover the entire mass range, its 14 TeV upgrade could give full discovery with
only $15\,\,\mbox{fb}^{-1}$.

\section{Conclusions}
\label{conclusions}
In this paper, we have studied the phenomenology of the four-site Higgsless 
model, which is an extension of the minimal three-site version (BESS model). 
The four-site model is a deconstructed theory based on the
$SU(2)_L\times SU(2)_1\times SU(2)_2\times U(1)_Y$ gauge symmetry. It predicts 
four charged ($W^\pm_{1,2}$) and two neutral ($Z_{1,2}$) extra gauge bosons. 
We have focused on the properties of the neutral gauge sector and on its 
discovery prospects at the Tevatron and the 7 TeV LHC in the dielectron 
channel during the next two years.

The phenomenology of the four-site model is controlled by only three free 
parameters beyond the SM ones: the two $Z_{1,2}$-boson masses and the 
$Z_2$-boson coupling to SM electrons. At fixed ratio 
$z=M_1/M_2\simeq M_{Z_1}/M_{Z_2}$, the model can be thus visualized in the 
($M_{Z_2}, \hat{a}^\Pe_{2L}$) plane. The parameter space has a twofold bound. 
On one side, unitarity requirement and $\mathcal{O}(x)$-approximation  
restrict the calculability of the model within the following mass range:
$350\le M_{Z_1}\le 1800 \GeV$ and $600\le M_{Z_2}\le 3000 \GeV$. On the other 
side, EWPT constrain the $\hat{a}^\Pe_{2L}$ coupling. Nevertheless, the 
allowed parameter space is sizeable. This represents the major novelty of the 
four-site model which, in contrast to common Higgsless theories, can solve 
the dichotomy between unitarity and EWPT bounds without imposing the extra 
vector bosons to be fermiophobic. As a consequence, the Drell-Yan process 
becomes a relevant channel for the direct search of extra gauge bosons at the 
LHC and the Tevatron.

We have first described main properties and shape of the new 
$Z_{1,2}$-bosons. We have shown their total widths and branching ratios into 
fermions and bosons. The results can be summarized in two points. First, even 
if $Z_{1,2}$ decay preferably into bosons, BRs into lepton pairs and boson 
pairs decaying in turn into leptons can compete, when looking at purely 
leptonic final states. For the four-site Higgsless model, the DY process is 
thus a strong discovery channel. Second, in contrast to common 
$Z^\prime$-bosons which purely decay into SM fermions and appear as narrow 
resonances, $Z_{1,2}$ can display a broad peaking structure. Their total 
width could be measured in large portions of the parameter space, offering 
one more challenge to the experiments. 
In order to show what signature might be expected, we have presented three 
sample scenarios corresponding to different points in the parameter space.
Here, the two $Z_{1,2}$-bosons could appear as resonances either quite distant
in mass or almost degenerate. If $M_{Z_1}\simeq M_{Z_2}$, two problems arise: 
their experimental separability or their collapse into just one peak. In this 
latter event, the four-site model would loose its characteristic 
multi-resonant structure becoming degenerate with common single-$\PZ^\prime$ 
models.

We have then computed the present direct limits on $Z_{1,2}$-boson masses and
coupling from recent searches performed by the D0 collaboration at the 
Tevatron with collected luminosity L=3.6 fb$^{-1}$. The outcome is that the
four-site model is weakly bounded: only a small corner around the minimum
mass, $M_{Z_2}\simeq$600 GeV, has been excluded up to now.  
 
The (almost) full parameter space is thus an open ground for the next two year
experiments at the Tevatron and the LHC. Owing to low signal acceptance 
($\simeq$20$\%$) and reduced energy, the Tevatron with project 
luminosity L=10 fb$^{-1}$ is not expected to play a major role in  
%go beyond 800 GeV for the 
the $Z_{1,2}$-boson discovery. Its exclusion bounds can however extend up to 
1 TeV scale. These rates get enhanced at the 7 TeV LHC which by the end of 
2011 should reach an integrated luminosity  L=1 fb$^{-1}$.
Already in the next few months, with a preliminary luminosity of 50 
pb$^{-1}$, the LHC can recover the present exclusion limits from Tevatron 
data. By collecting additional luminosity, the 7 TeV LHC should then be able 
to discover new physics predicted by the four-site Higgsless model up to 
1400 GeV. The exclusion limits could cover the whole mass spectrum, depending 
on model parameters. 

Finally, looking at the future, we have shown how the project 14 TeV LHC with 
L=10 fb$^{-1}$ could sensibly extend the four-site physics search to regions
of the parameter space with small $Z_{1,2}$-boson couplings to ordinary matter.

\noindent
{\bf{Note added in proof.}} During completion of this work, we became aware
of latest results obtained by the D0-collaboration on the search for a heavy 
neutral gauge boson in the dielectron channel at the Tevatron with 
5.4 fb$^{-1}$, published in Ref.\cite{D0-2010}. The new 95$\%$ C.L. upper
limit on the $Z^\prime$ cross section can shift the $Z_{1,2}$-boson mass 
bound up to $M_{Z_2}\simeq$800 GeV in a small portion of the parameter space
($z$=0.8 and maximal $\hat{a}^\Pe_{2L}$ coupling). 
The projection for the next two year running at the Tevatron is basically 
unchanged.

\acknowledgments
LF would like to thank the School of Physics and Astronomy of the University 
of Southampton for hospitality. We are very grateful to L. Basso and G. M. 
Pruna for comparisons. We also acknowledge S. King, A. Belyaev and 
C. H. Shepherd-Themistocleous for valuable discussions.

\appendix

\section{Gauge boson trilinear couplings}
\label{appendixA}
The general form for the trilinear vertex is the following:
\be
\mathcal{L}_3=i \sum_{R,P,Q} a_{RPQ}(R_{\mu\nu}^+ P_\mu^- Q_\nu-P_{\mu\nu}^- R_\mu^+ Q_\nu+ Q_{\mu\nu} P_\mu^+ R_\nu^-)\equiv i \sum_{R,P,Q} a_{RPQ} O_{RPQ}
\ee
where $R$, $P= W$, $ W_1$, $W_2$ and $Q=\gamma$, $Z$, $ Z_1$, $Z_2$
and
\begin{equation}
a_{WW\gamma}=a_{W_1W_1\gamma}=a_{W_2W_2\gamma}=e
\ee
\be
a_{WWZ}=\gt c_{\tilde\theta} \left[1-\frac{x^2}{2c^2_{\tilde\theta}}\left(1-2s^4_{\tilde\theta}\right)\right]
\end{equation}
\begin{equation}
a_{WWZ_1}=-\frac{\gt x(1-z^4)}{2\sqrt{2}}\label{AA4}
\end{equation}
\begin{equation}
a_{WW_1Z}=-\frac{\gt x(1-z^4)}{2\sqrt{2}c_{\tilde\theta}}\
\end{equation}
\begin{equation}
a_{WW_1Z_1}=-\frac{\gt}{2}\left[1-\frac{x^2}{4c^2_{\tilde\theta}}\left(1-(4-z^2)+\frac{z^2}{1-z^2}+\frac{z^2(1+2z^2)c_{2{\tilde\theta}}}{1-z^2}\right)\right]
\end{equation}
\begin{equation}
a_{WW_1Z_2}=\frac{ \gt z^2}{2}\left[1+\frac{x^2z^2}{4 c^2_{\tilde\theta}}\left(2c_{2{\tilde\theta}}-(4-s^2_{\tilde\theta})z^2-\frac{2z^2s^2
_{\tilde\theta}}{1-z^2}\right)\right]\label{AA7}
\end{equation}
\begin{equation}
a_{WW_2Z_1}=-\frac{z^2 \gt}{2}\left[1+\frac{x^2}{4 c^2_{\tilde\theta}}\left(2z^2-4z^4c^2_{\tilde\theta}+\left(1-\frac{2z^4}{1-z^2}s^2_{\tilde\theta}\right)\right)\right]
%\left(\right) \left(1+\frac{\gt^2}{4 g_1^2 \cos^2{\tilde\theta}}\left(\right)\right) \frac{\gt  {g_1^2}
\end{equation}
\bea
a_{WW_2Z_2}&=&-\frac{\gt}{2}\left[1+\frac{x^2}{4c^2_{\tilde\theta}}\left(\frac{1+3z^2(z^2-1+z^4)-s^2_{\tilde\theta}(1-3z^2+4z^4+4z^6)}{1-z^2}\right)\right]
\eea
\bea
a_{W_1W_1Z}&=&-\frac{\gt c_{2{\tilde\theta}}}{2c_{\tilde\theta}}\left[1+\frac{x^2}{4c^2_{\tilde\theta}c_{2{\tilde\theta}}}\left(\frac{3-5z^2-3z^4+z^6}{1-z^2}+\right.\right.\nn\\
&&\left. \left. +s^2_{\tilde\theta}\left(2z^4+\frac{4z^2}{1-z^2}-5-2c_{2{\tilde\theta}}-c_{4{\tilde\theta}}\right)\right)\right]
\eea
\begin{equation}
a_{W_1W_1Z_1}=-\frac{\gt}{x\sqrt{2}}\left[1-\frac{x^2}{4c_{\tilde\theta}}\left(2c^2_{\tilde\theta}+1\right)\right]
\end{equation}
\begin{equation}
a_{W_1W_1Z_2}=-\frac{\gt xz^2(2-z^2+z^2t^2_{\tilde\theta})}{2\sqrt{2}(1-z^2)}
\end{equation}
\begin{equation}
a_{W_1W_2Z_2}=\frac{\gt}{\sqrt{2}x}\left[1-\frac{x^2}{4  c^2_{\tilde\theta}}\left(z^4+(1+z^4)c^2_{\tilde\theta}\right)\right]
\end{equation}
\begin{equation}
a_{W_1W_2Z_1}=\frac{\gt x z^2\left(z^2-3+\frac{1}{c^2_{\tilde\theta}}\right)}{2\sqrt{2}(1-z^2)}
\end{equation}
\bea
a_{W_1W_2Z}&=&\frac{z^2\gt}{2c_{\tilde\theta}}\left[1+\frac{x^2}{4c^2_{\tilde\theta}}\left(2z^2(1-2z^2)+s^2_{\tilde\theta}(1+z^4+2c_{2{\tilde\theta}})\right)\right]
\eea
\bea
\!\!\!\!\!\!\!\!\!a_{W_2W_2Z}&=&\frac{\gt c_{2{\tilde\theta}}}{2c^2_{\tilde\theta}}
\left[1+\frac{x^2}{4g_1^2c^2_{\tilde\theta}c_{2{\tilde\theta}}}
\left(4z^2-3-\frac{4z^6c^2_{\tilde\theta}}{1-z^2}+(1-z^2)^2s^2_{\tilde\theta}-(2c_{2{\tilde\theta}}+c_{4{\tilde\theta}})s^2_{\tilde\theta}
\right)\right]
\eea
\begin{equation}
a_{W_2W_2Z_1}=\frac{\gt }{\sqrt{2}x}\left[1-\frac{x^2}{4c^2_{\tilde\theta}}\left(1+2z^4c^2_{\tilde\theta}\right)\right]
\end{equation}
\begin{equation}
a_{W_2W_2Z_2}=\frac{\gt xz^4(1+2c_{2{\tilde\theta}})}{2\sqrt{2}(1-z^2)c^2_{\tilde\theta}}
\end{equation}
All couplings not listed are zero to the order  $x^2$.

\addcontentsline{toc}{chapter}{References}
%\bibliographystyle{h-physrev4}  
%\bibliography{bib_last}

\hyphenation{Post-Script Sprin-ger}

\end{document}